\documentclass[longbibliography,aps,prapplied,reprint,superscriptaddress]{revtex4-2}
\pdfoutput=1
\usepackage{blindtext}
\usepackage{natbib}
\usepackage{wrapfig}  
\usepackage{mathrsfs,amsmath,amssymb,color,bm,mathtools}
\usepackage[makeroom]{cancel}
\usepackage{setspace}
\usepackage{booktabs}

\usepackage[english]{babel}

\usepackage{bbold}
\usepackage{siunitx}

\DeclareSymbolFont{ntxletters}{OML}{ntxmi}{m}{it}  
\DeclareMathSymbol{\tau}{\mathord}{ntxletters}{28}  

\newcommand{\plusequals}{\mathrel{+}=}

\newcommand{\subfiglabel}[1]{(#1)}
\newcommand{\figref}[1]{Fig.~\ref{#1}}
\newcommand{\subfigref}[2]{\figref{#1}\subfiglabel{#2}}

\newcommand{\appref}[1]{Appendix~\ref{#1}}
\newcommand{\eqnref}[1]{Eq.~\eqref{#1}}

\newcommand{\tens}{T} 
\newcommand{\tchar}{\tens^*}

\providecommand{\red}[1]{{\color{red}{#1}\color{black}}}
\providecommand{\blue}[1]{{\color{blue}{#1}\color{black}}}

\newcommand{\papertitle}{Stopping and reversing sound via dynamic
  dispersion tuning in a phononic metamaterial}

\makeatletter
\DeclareRobustCommand{\iscircle}{\mathord{\mathpalette\is@circle\relax}}
\newcommand\is@circle[2]{%
	\begingroup
	\sbox\z@{\raisebox{\depth}{$\m@th#1\circ$}}%
	\sbox\tw@{$#1\square$}%
	\resizebox{!}{\ht\tw@}{\usebox{\z@}}%
	\endgroup
}
\makeatother

\begin{document}
	\title{\papertitle}
	\author{Pragalv Karki}
	\email{pragalvk@uoregon.edu}
	\affiliation{Department of Physics and Institute for Fundamental Science, University of Oregon, Eugene, OR 97403, USA}
	\author{Jayson Paulose}
       	\email{jpaulose@uoregon.edu}
	\affiliation{Department of Physics and Institute for Fundamental Science, University of Oregon, Eugene, OR 97403, USA}
	\affiliation{Material Science Institute, University of Oregon, Eugene, OR 97403, USA}
	
	\begin{abstract}
          Slowing down, stopping, and reversing a signal is a core functionality for
          information processing. Here, we show that this functionality can be realized by
          tuning the dispersion of a periodic system through a dispersionless, or flat, band. Specifically, we
          propose a phononic metamaterial based on plate resonators, in
          which the phonon band dispersion can be modified from an acoustic-like to an optical
          character by modulating a uniform prestress. The switch is enabled
          by the change in sign of an effective coupling between fundamental modes, which
          generically leads to a nearly dispersion-free band at the transition point. We
          demonstrate how adiabatic tuning of the band dispersion can immobilize and
          reverse the propagation of a sound pulse in simulations of a one-dimensional
          resonator chain. Our study relies on the basic principles of thin-plate
          elasticity independently of any specific material, making our results applicable
          across varied length scales and experimental platforms. More broadly, our
          approach could be replicated for signal manipulation in photonic metamaterials
          and electronic heterostructures.          
	\end{abstract}
	
\maketitle

\section{Introduction}

The tunability of sound transport properties after fabrication is a prominent
feature underlying the appeal of phononic metamaterials~\cite{Wang2020}. In
periodic structures, tunability can be achieved by modifying the band structure
of vibrational excitations, which determines both the frequency ranges of sound
insulation (via bandgaps) and the group velocity of sound propagation (via the
frequency-momentum relationship or dispersion relation). Metamaterials with
tunable phononic bands have been proposed which use modulation methods as varied
as buckling~\cite{Wang2014,Bertoldi2017a}, large structural
deformations~\cite{Babaee2016,Hedayatrasa2016,Pal2016},
electrical~\cite{Casadei2012,Cha2018,Yi2019} and optical~\cite{Swinteck2014}
actuation, and prestress
modulation~\cite{Feng2012,BARNWELL201723,Krushynska2018,Pal2018a,Li2020a}. While
most of these proposals have targeted the tuning of bandgaps,
several works~\cite{Casadei2012,Pal2016,Cha2018,Pal2018a} have highlighted the ability
to change group velocities by tuning the dispersion relation as a promising
direction for adapting a metamaterial to different static conditions.
However, the dynamic control of dispersion remains unexploited as a mechanism
for signal manipulation. 

\begin{figure}[tb]
	\noindent\includegraphics[width=\columnwidth]{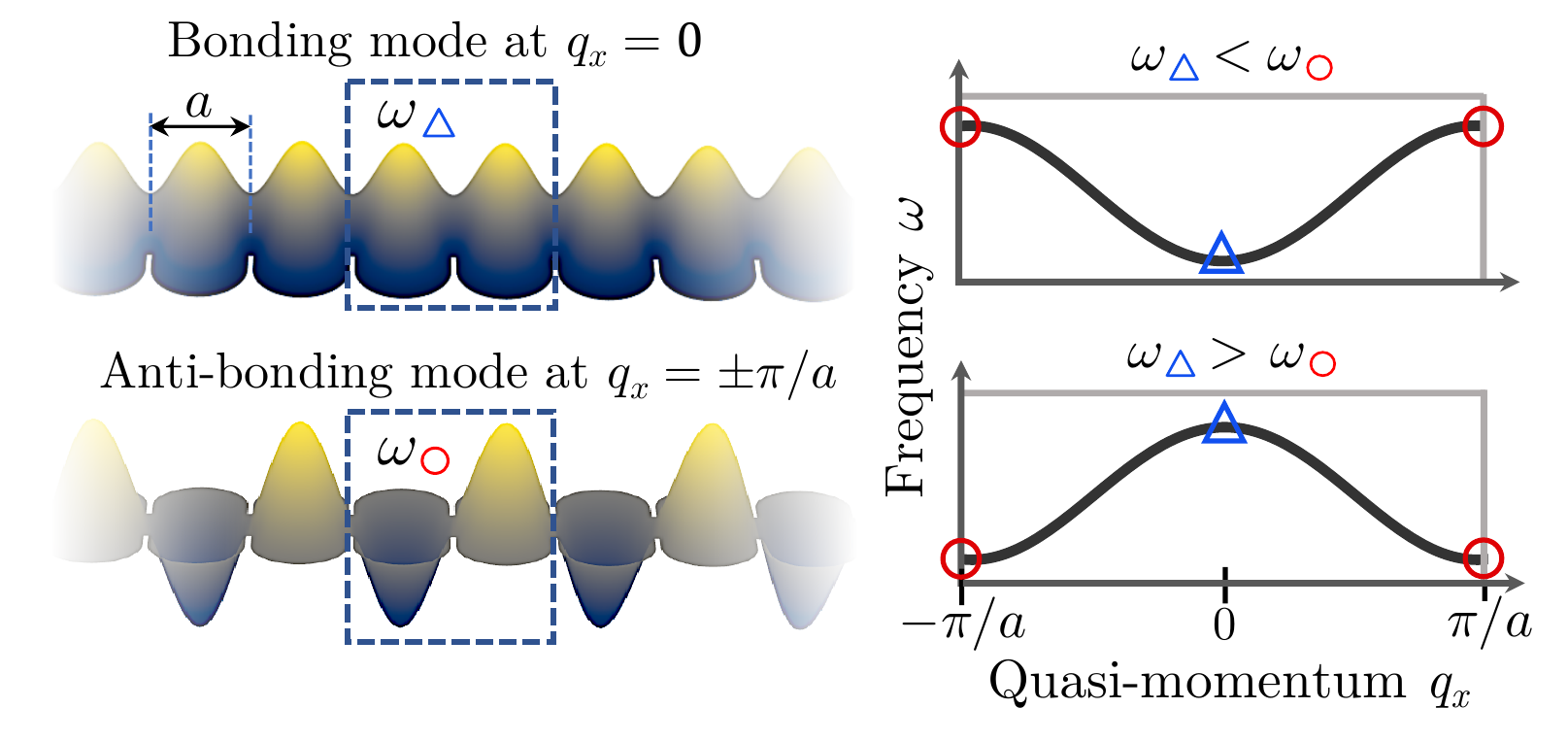}   
	\caption{   
	\label{illustration}
        Relationship between paired-mode frequencies and dispersion, illustrated
        for a periodic system with lattice constant $a$. \emph{Left,} The Bloch
        mode of the lowest band at $q_x=0$ is composed of a chain of
        in-phase or bonding pairs of fundamental excitations (top), whereas the
        mode at $q_x=\pm\pi/a$ is composed of out-of-phase or antibonding pairs
        (bottom). \emph{Right,} Schematic of the different band dispersion
        relations when the frequency of the bonding pair
        $\omega_{\bm{\blue{\triangle}}}$ is lower (top) or higher
        (bottom) than the antibonding pair frequency
        $\omega_{\red{\iscircle}}$.}
\end{figure}

In this work, we demonstrate how to stop and reverse signals in a
tunable metamaterial by dynamically changing the dispersion character of an
entire band during signal propagation.
Specifically, we describe a physical mechanism to flip the sign of the group
velocity across all quasimomenta (i.e., wavevectors associated with the excitations of
the periodic lattice), thereby reversing the propagation direction of
wave pulses (\figref{illustration}). A unique aspect of our approach is that
the group velocity vanishes 
throughout the band at the point of sign switching, giving rise to a flat phononic
band which can trap a signal for subsequent release when required. Using full-wave finite element simulations, we show how
adiabatic tuning of the dispersion allows us to store and reverse a sound pulse
in a waveguide---a functionality which has potential applications in acoustic
sensing~\cite{Fatemi1998,Fu2017}, signal processing~\cite{1978tap}, and
computation~\cite{Li-Feng,Zangeneh-Nejad2018,Wang}.

\begin{figure}[t]
	\noindent\includegraphics[width=\columnwidth]{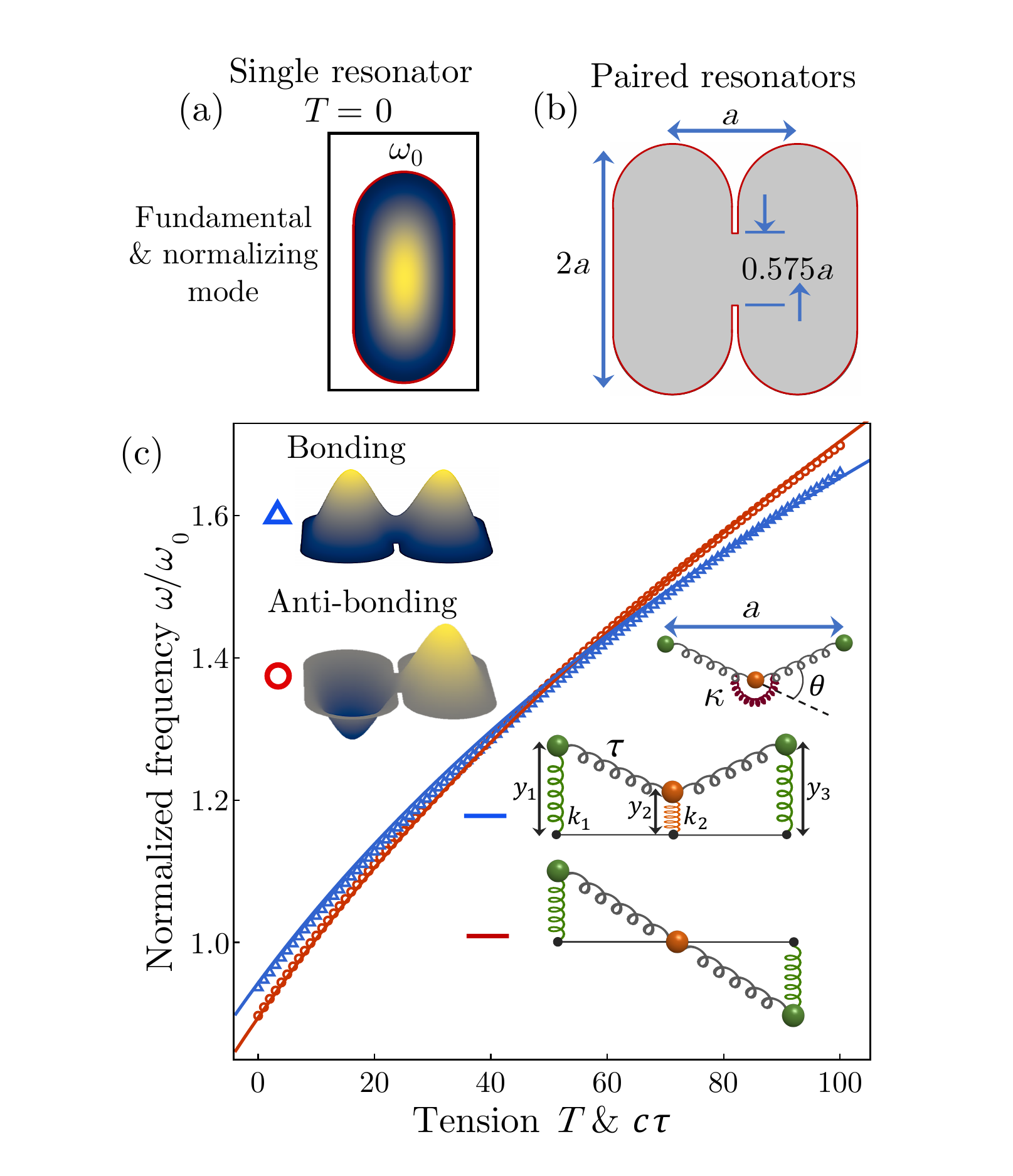}   
	\caption{   
          \label{ModeCrossing}
          (a) Geometry of a single resonator supported by a rigid boundary (red edge)
          and free to vibrate in the interior. Color intensity represents the
          displacement field of the fundamental mode at zero tension, whose
          frequency $\omega_0$ is used as the normalizing frequency throughout
          the study. (b) Geometry of a resonator pair including the junction.
          This specific junction geometry was chosen to maximize mode separation at
          $T=0$, but the effect persists for other narrow junctions.
          (c) Frequency of the two lowest eigenmodes of the
          resonator pair as a function of rescaled tension. Symbols are from
          finite-element simulations, and distinguish the bonding character of the
          corresponding eigenmode. Left inset shows
          numerically-determined eigenmodes at $T=0$. Solid lines show lowest two
          eigenfrequencies of the discrete dynamical matrix,
          \eqnref{DynamicalMatrix-3M}. Dimensionless parameters $T$ (continuum
          model) and $\tau$ (discrete model) are related via $T = c\tau$, where
          $c=47.5$ is a geometry-dependent linear mapping constant.
          Right inset shows tensile and torsional springs in the
          discrete model, and examples of bonding (blue) and antibonding (red)
          modes. }
\end{figure}

We accomplish the desired change in dispersion by manipulating the coupling
between adjacent degrees of freedom in a periodic structure. In our design,
the relevant degrees of freedom are the fundamental (i.e., lowest-frequency)
transverse vibrational modes of free-standing thin-plate mechanical 
resonators supported by a rigid frame. However, the underlying physical principle is independent of the specific type
of excitation, as illustrated in \figref{illustration} for the lowest excitation
band of coupled modes on an infinite periodic chain.
In a tight-binding description of coupled excitations, the Bloch state at the band
center (quasimomentum $q_x=0$) is constructed from eigenmodes in a ``bonding''
configuration (adjacent eigenmodes are in-phase), while the state at the band
edge ($q_x=\pm \pi/a$ where $a$ is the lattice constant) is an assembly of
out-of-phase or ``antibonding'' pairs. If the bonding
state for a pair of building blocks is at a lower frequency than the antibonding
one, the dispersion relation 
must increase from the band center to the band edge. In contrast, if the
antibonding configuration has a lower frequency, the dispersion relation is a
decreasing function of the quasimomentum magnitude. Therefore, flipping the
bonding character of pairwise couplings across a periodic structure can reverse
the group velocity (the slope of the dispersion relation) throughout the band. Below, we show that the
mechanics of thin plates under tension enables precisely such a reversal
(\figref{ModeCrossing}). However, the approach could be replicated in other
wave systems where the bonding character of paired degrees of freedom can be
controlled, such as photonic crystals~\cite{Caselli2012} or electronic
heterostructures~\cite{Yakimov2009,PhysRevA.91.023409}.

\section{Theoretical framework}

\subsection{Model of coupled plate resonators} 

We model the out-of-plane vibrations of suspended resonators with a defined edge geometry using the
partial differential equation for the transverse displacement field $u(x,y)$ of an elastic plate with mass per unit area $\rho$,
bending modulus $D$, subjected to a uniform
in-plane tension $T^\prime$ and clamped to a rigid plane curve along its edges~\cite{timoshenko1959theory}:
\begin{align} \label{full-pde}
	\begin{split}	
		\rho \frac{\partial^2 u}{\partial t^2} +
                D\nabla ^4 u - T^\prime\nabla^2 u = 0 \hspace{0.5cm} \text{on domain}, \\ ~~{}
		u=\nabla u=0 \hspace{0.5cm} \text{on boundary}.
	\end{split}
\end{align} 
While the bending modulus and density are materials
properties, the tension is an externally-imposed stress which can be tuned
through external manipulation (e.g. via laser heating~\cite{Blaikie2019} or
electrostatic gating~\cite{Cha2018}).

To perform our analysis without making explicit choices for the
physical dimensions and materials parameters,  we non-dimensionalize the
continuum plate equation \eqnref{full-pde} by performing a change of 
variables~\cite{Yosibash2005} using the lattice constant $a$ and the time
scale $\sqrt{\rho a^4/D}$ as the length and time units. Upon defining $\bar{x}=x/a$, $\bar{y}=y/a$,  
$\bar{t}=t\sqrt{D/(\rho a^4)}$, \eqnref{full-pde} yields
\begin{align} \label{non-dim-pde}
\begin{split}	
\frac{\partial^2 u}{\partial \bar{t}^2} 
+\bar\nabla^4  u  -  T \bar\nabla^2 u = 0 \hspace{0.5cm} \text{on domain}, \\ ~~{}
		u=\bar\nabla u=0 \hspace{0.5cm} \text{on boundary}. 
\end{split}
\end{align} 

Once the boundary geometry is specified, \eqnref{non-dim-pde} shows
that the non-dimensionalized dynamics depend on a single dimensionless
parameter---the rescaled tension $\tens \equiv T^\prime a^2/D$, which serves as
the tunable physical quantity in our study.
In the remainder of this
manuscript, we drop the bar for clarity; the variables $x$, $y$, $t$ and the gradient
operator $\nabla$ are understood to refer to the rescaled coordinates.

For a particular resonator geometry, oscillatory solutions to
\eqnref{non-dim-pde} can be expanded in terms of normal modes
$u_i(x,y)e^{-i \omega_i t}$ indexed by the variable $i$, where the functions
$u_i(x,y)$ and oscillation frequencies $\omega_i$ solve the eigenvalue problem
\begin{equation}
  \label{eq:eigenvalue}
  \nabla^4 u_i - T \nabla^2 u_i = \omega_i^2 u_i
\end{equation}
together with the boundary conditions. Normal mode
displacements and frequencies are computed using finite-element
analysis, see \appref{appendixFiniteElementAnalysis} for details.  
We use the lowest-frequency, or fundamental, mode of a single resonator with
frequency $\omega_0$
(\figref{ModeCrossing}(a)) as the basic degree of freedom in our system and
consider the collective modes that arise upon coupling fundamental modes across
multiple resonators through junctions as shown in
\figref{ModeCrossing}(b).

For narrow junctions, the lowest two eigenmodes of a pair of resonators can be
identified with a bonding and an antibonding configuration of the
fundamental modes of the individual resonators
(\figref{ModeCrossing}(c)). In the absence of a bending stiffness,
\eqnref{eq:eigenvalue} reduces to a Laplacian eigenfunction problem, for which the
maximum principle dictates that the eigenfunction with the lowest eigenvalue
must be of fixed sign over the domain. As a result, the bonding mode is
guaranteed to be lower in frequency than the antibonding mode in the $D\to 0$,
or $\tens \to \infty$, limit. However, the maximum principle does not hold for
the \emph{biharmonic} eigenfunction problem obtained in the $\tens \to 0$ limit
of \eqnref{eq:eigenvalue}, for which domains with non-convex boundaries have
been found to favor lowest-frequency eigenfunctions with sign changes
within the domain~\cite{Sweers2001,brown1999accurate}.
Therefore, we expect the antibonding mode to be at lower frequency when
the external tension is set to zero, but to switch to higher frequency relative
to the bonding mode at large tensions. Numerical solutions of the lowest two
eigenmodes of the resonator pair, obtained via finite-element analysis (Appendix \ref{appendixFiniteElementAnalysis}),
confirm this expectation (\figref{ModeCrossing}(c)). The lowest-frequency mode
switches from antibonding to bonding type at a geometry-dependent threshold tension
$\tchar \approx 45$, at which the two lowest normal mode frequencies coincide
to signify a degeneracy of the antibonding and bonding modes.

Since the crossing behavior fundamentally arises from the competition between
bending and tension in thin-plate mechanics, it is not restricted to our particular
choice of single-resonator geometry in \figref{ModeCrossing}. While details such
as the magnitude of the threshold tension are geometry-dependent, the basic
mechanism is generic to a wide range of resonator geometries when
connected by a narrow junction. Furthermore, while we have used the clamped
boundary condition to capture the typical edge constraint for micromechanical
resonators mounted on semiconductor substrates~\cite{Yu2012}, the switch in
bonding character also occurs for simply-supported edges (Dirichlet boundary
conditions), as we verify in Appendix \ref{appendixBoundaryCondition}.

\subsection{Minimal model of mode-crossing mechanism} 

The balance between bending and tension which drives the eigenmode crossing can be captured in
a simpler discrete model of coupled harmonic oscillators.
The fundamental mode of an isolated resonator
is modeled as a harmonic degree of freedom $y$ confined to the vertical direction, with
a unit mass and spring constant $\tilde{k}_1$. Building on past
approaches~\cite{Matlack2018}, we then attempt to build
the normal modes of coupled resonators by incorporating couplings among
fundamental modes on adjacent oscillators, using e.g. a horizontal spring under
tension. However, according to the von~Neumann-Wigner theorem, coupling two
degrees of freedom would generically create an \emph{avoided} crossing
of eigenvalues of the coupled system upon varying the coupling strength,
in contrast to the observed behavior in \figref{ModeCrossing}(c).

The key to obtaining the correct crossing behavior is to
incorporate the plate deformation at the junction into the reduced description
as an additional degree of freedom---specifically, a unit mass on a vertical
spring with stiffness $\tilde{k}_2 > \tilde{k}_1$. This mass is coupled to the resonator degrees
of freedom through tensed and torsional springs, as shown schematically in the
right inset to \subfigref{ModeCrossing}{c}.
The coupling of the resonator modes due to
tension is encoded in harmonic springs connecting each mode mass to the junction
mass, which are prestressed with a tensile force $\tilde{\tau}$. These contribute a
potential energy $U_s = \tilde{\tau}((y_1-y_2)^2 + (y_2-y_3)^2)/a$ to vertical
displacements. The bending stiffness penalizes geometric curvature at the
junction; we include this effect by defining a torsional spring which favors
collinearity of the two tensile springs with associated harmonic energy
$U_b = \tilde{\kappa}(1-\cos\theta) \approx 2\tilde{\kappa} ( y_1 -
2y_2+y_3)^2/a^2$. By choosing $\omega_0^{-1}$ and $a$ as the time and length
units respectively, we obtain a discrete model with four dimensionless
parameters $k_1$, $k_2$, $\tau$, and $\kappa$  (the absence of the tilde
indicates non-dimensionalized quantities), Appendix \ref{appendixDiscreteNonDimensionalization}. The 
non-dimensionalized stiffness matrix 
describing the dynamics of the three harmonic degrees of freedom is
\begin{equation}
	\mathbf{K} = {\begin{pmatrix} k_1+\tau+\kappa & -\tau-2\kappa & \kappa \\ -\tau-2\kappa & k_2+2\tau+4\kappa & -\tau-2\kappa \\ \kappa & -\tau-2\kappa & k_1+\tau+\kappa \end{pmatrix}} 
	\label{DynamicalMatrix-3M}
.
\end{equation}

The two lowest eigenfrequencies of the stiffness matrix correspond to the coupled modes
that arise from weak-coupling of the fundamental modes (Appendix \ref{appendix3SiteModel}); the third mode
 is considerably higher in frequency and is not relevant to our analysis. When $\tau =0$, the
antibonding configuration has a lower frequency because the torsional spring
remains undistorted (\figref{ModeCrossing}(c), lower right). Upon increasing
$\tau$ with other parameters kept fixed, the bonding configuration becomes
increasingly favored because it costs lower tensile energy, and the two modes
become degenerate at $\tau = \sqrt{(3\kappa/2)^2 + \kappa(k_2-k_1)} -
3\kappa/2$. If a linear relationship is assumed between the dimensionless
tensions $\tau$ and $T$, the minimal model with four fit parameters
quantitatively captures the evolution of the normal mode 
frequencies from the continuum model (compare symbols to solid lines
in \figref{ModeCrossing}(c)). Details of the fitting procedure are provided in
\appref{appendixFitProcedure}. 

\begin{figure}[htb]
	\noindent\includegraphics[width=\columnwidth]{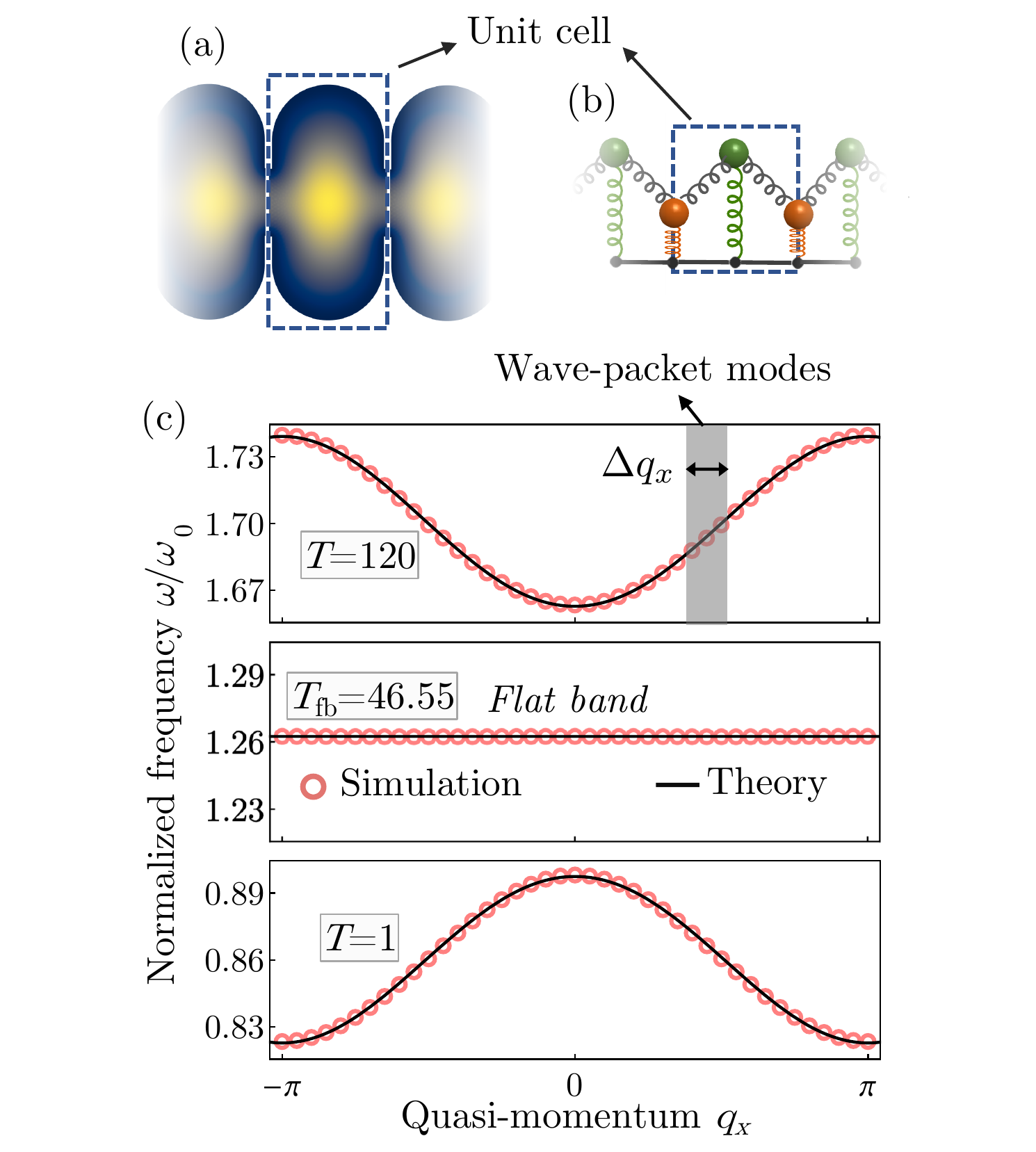}  
	\caption{   
          \label{Band-diagram}
          Dispersion relation of the fundamental band. (a) Unit cell of the
          continuum model, with intensity variation representing the
          displacement of the Bloch eigenmode at $q_x=0$ from finite-element
          calculations. (b) Unit cell of the discrete model. (c) Band structures
          of the continuum model at three values of rescaled tension (symbols),
          compared to bands computed from the discrete model (solid lines). Fit
          parameters $\tau$ and $k_1$ for different tensions $T$ are: $T=1$
          $\rightarrow$ $\{\tau,k_1\}\approx\{0.3195,0.0381\}$, $T=46.55\equiv T_\text{fb}$
          $\rightarrow$ $\{\tau,k_1\}\approx\{0.5217,0.5502\}$, and $T=120$
          $\rightarrow$ $\{\tau,k_1\}\approx\{0.7677,1.4888\}$; see
          \appref{appendixFitProcedure} for details of fitting procedure. Shaded region
          shows the quasimomentum range used to 
          to create the Gaussian wavepacket for the dynamical simulation.
        }
\end{figure}

\section{Results}

\subsection{Dispersion relation and flat band} 

To illustrate the consequences of the switch in parity of the lowest-frequency
pair eigenmode on sound transport, we compute the mode spectrum for an infinite
1D chain of coupled resonators as a function of the quasimomentum $q_x$ which
indexes the Bloch eigenfunctions $u_{q_x}(x,y) = e^{iq_xx} \phi_{q_x}(x,y)$. For the
continuum system, whose unit cell is depicted in \figref{Band-diagram}(a), the
numerically-determined spectrum (Appendix \ref{appendixFiniteElementAnalysis}) consists of infinitely many bands within the
Brillouin zone $-\pi/a < q_x < \pi/a$, but the lowest band is built
primarily from the fundamental modes of the individual resonators. The
dispersion relation $\omega(q_x)$ for this band is captured in the reduced
description with a two-mass unit cell shown in \figref{Band-diagram}(b). The
Fourier-transformed stiffness matrix obtained by upgrading
\eqnref{DynamicalMatrix-3M} to a periodic chain is
\begin{equation}
\mathcal{\mathbf{K}}(q_x)= \begin{pmatrix} a+b\cos q_x & -c(1+e^{-iq_x}) \\ -c(1+e^{iq_x}) & d \end{pmatrix}
\label{DynamicalMatrix-1D}
,
\end{equation}
where $a=k_1+2\tau+2\kappa$, $b=2\kappa$, $c=\tau+2\kappa$, and
$d=k_2+2\tau+4\kappa$. The frequency bands are then solved via
$|\mathcal{\mathbf{K}}(q_x)-\omega(q_x)^2 \mathbf{I}|=0$.

Dispersion relations for the lowest band, computed using both the continuum and
the discrete descriptions, are shown in \figref{Band-diagram}(c) for three
values of the globally-applied plate tension. Changing free parameters $\tau$ and $k_1$ 
in the discrete model effectively captures the changing of $T$ in the continuum 
model (Appendix \ref{appendix3SiteModel}). We find that the band  changes
from optical type (frequency decreasing with quasimomentum) to acoustic type
(frequency increasing with quasimomentum) as the tension is increased, in line
with our expectation (\figref{illustration}). At a special
value of the rescaled tension, the band becomes nearly dispersion-free, or flat.
In the discrete model, we analytically
establish the existence of a band with
$\partial \omega/\partial q_x = 0$ throughout the Brillouin zone when the condition
\begin{equation}
  \label{eq:flatbandtension}
  \tau = \sqrt{(2\kappa)^2 + \kappa(k_2-k_1)} - 2\kappa
\end{equation}
is satisfied (see Appendix \ref{appendixFlatBandDerivation} for details), so a tension
value leading to a perfectly flat band can always be found provided $k_2 > k_1$. In the
continuum model, the dispersion does not completely vanish; however, the bandwidth is
limited to $10^{-3}\%$ of the mean band frequency. This minute deviation from a
perfectly flat band can be reproduced in the discrete model by adding one more
torsional spring to the unit cell, centered on the resonator mass (Appendix \ref{appendixFlatnessDeviation}).

\begin{figure*}[tb]
	\noindent\includegraphics[width=\textwidth]{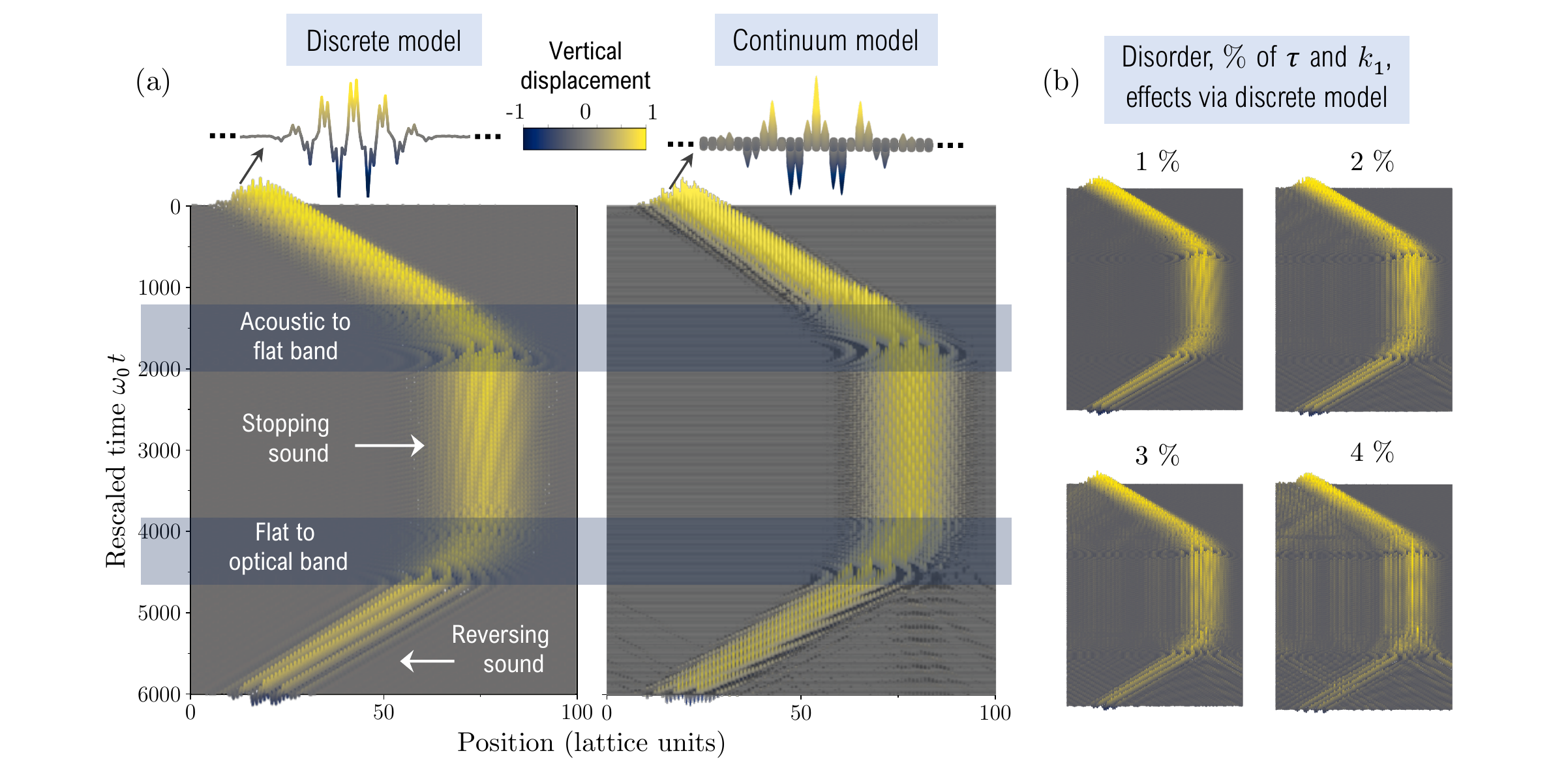}  
	\caption{   
		\label{wave-packet}
		(a) Sound pulse reversal through dynamic dispersal tuning. Color map shows
		the vertical displacement evolution of Gaussian wavepackets in dynamical 
		simulations of a linear chain of coupled resonators, for both the discrete
		(left) and the continuum (right) models. 
		Initial pulses are shown for both systems at the top. Time advances 
		from top to bottom; horizontal axis is position along the chain. Shaded time intervals
		are periods during which the dispersion was varied
		along a linear ramp connecting initial to final parameter values as described in
		the text. (b) Disorder effects on wave packet manipulation with 
		a quenched disorder of 1\%--4\% of $\tau$ and $k_1$ throughout the chain.
		Dynamically slowing and stopping the pulse is robust up to $3\%$ disorder 
		and the reversal is robust up to $2\%$. At higher disorder levels,
		reversal is partially successful for this set of parameters, but a considerable 
		amount of energy is lost to spurious modes.
	}
\end{figure*}

We note here that the lowest frequency in the
periodic system does not go to zero at $q_x =0$ or $q_x = \pi$, i.e. a
low-frequency bandgap exists in the resonator system. This bandgap is a
consequence of the finite frequency of the fundamental mode. When the tension is
set to zero, the vibrational frequency of a continuum displacement field scales
as $\sqrt{D/(\rho l^4)}$ where $l$ is the shortest length scale of variations of the
displacement field. For a single resonator, $l$ is at most some fraction of $a$
and the fundamental frequency is of order $\sqrt{D/(\rho a^4)}$ with a prefactor
larger than one [numerically, we find $\omega_0 \approx 27.3\sqrt{D/(\rho a^4)}$
in physical units,
consistent with the apparent variation of the displacement field which happens
over roughly a third of the resonator width in Figure 2(a)]. For an infinite 1D
chain of resonators, the lowest frequency remains of order $\omega_0$ because
the narrow junctions only weakly perturb the adjacent modes and the displacement
fields are still restricted by the resonator width $a$. If we were to widen the
junctions, the lowest frequency of the band would drop, but it would still be
bounded from below by the resonator height, $l \lesssim 2a \Rightarrow \omega
\gtrsim 4\sqrt{D/(\rho a^2)}$ and a finite bandgap
is always expected. By contrast, for a two-dimensional resonator array we could
nearly eliminate the bandgap by widening the coupling junctions in both
directions. In that case, the vibrational response of the lowest-frequency Bloch
mode would approach that of a square plate of infinite dimension, which truly
goes to zero as $l \to \infty$.

\subsection{Slowing, stopping, and reversing a sound pulse} 

\begin{figure*}[tb]
	\noindent\includegraphics[width=\textwidth]{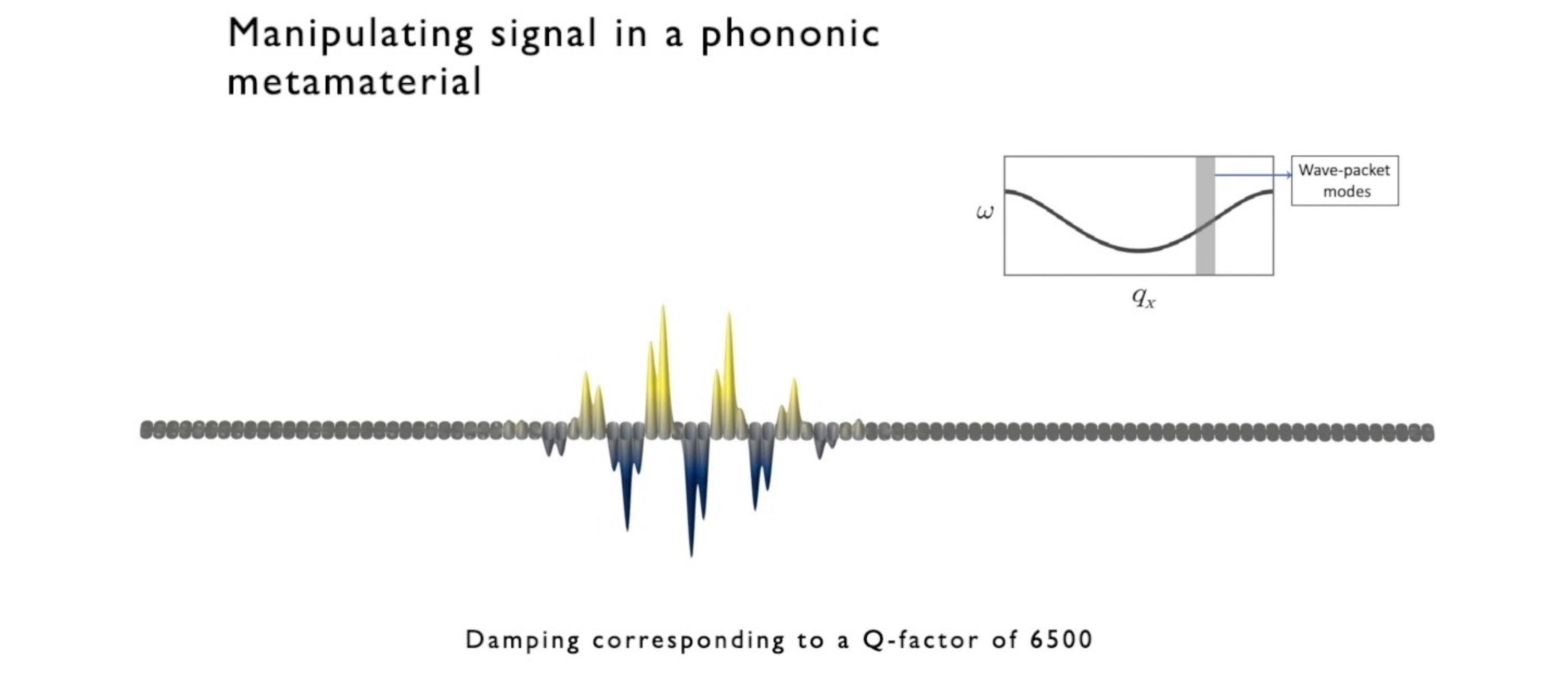}   
	{\small 
		Video 1. Full-wave finite element simulation which shows the stopping and reversing of a 
		sound pulse through dynamic dispersion tuning. The image above
		shows a single frame from the video.
		The video starts with a wave-packet prepared at the left end of the 1D array 
		with a positive group velocity moving to the right.
		The wave-packet is stopped and
		reversed by adiabatically tuning the tension as described in the
		text. The corresponding change in the dispersion is displayed in
		the inset on the upper right. The wavevector components comprising the
		wave-packet are highlighted in gray.
	}
\end{figure*}

As a prototypical example of manipulating a signal by dynamically tuning the
dispersion through a flat band, we consider the evolution of a Gaussian wavepacket under uniform
adiabatic modulation of the global tension in a resonator array.
We performed full-wave simulations of the continuum model via a finite-element method, and classical dynamics simulations of the discrete mass-spring
model via the velocity Verlet algorithm (Appendix \ref{appendixDynamicalSimulation}).
For both models, a 1D array of 100 unit cells was initialized with a wavepacket
built from the band corresponding to $T=120$ in \figref{wave-packet}. The Bloch eigenfunctions
$\boldsymbol{\phi}_{q_x}$ were used
to construct a sound pulse using the equation
\begin{equation}
\boldsymbol{u}(n,t)= \sum_{q_x} \boldsymbol{\phi}_{q_x} e^{{i(q_{x} n - \omega(q_x) t)}} e^{-\left(\frac{q_x-q_0}{\Delta q_x}\right)^2},
\label{wave-packet-eq}
\end{equation}
where $\boldsymbol{u}(n,t)$ is the vector of displacements of the $n$th unit
cell at time $t$, and the sum is over all allowed discrete quasimomenta $q_x$.
In the discrete model, $\boldsymbol{\phi}_{q_x}$ and $\boldsymbol{u}$ are 2-vectors,
while in the continuum model their length is determined by the mesh size of the
finite-element discretization.
The pulse was initialized at $t=0$ with plane-wave weights centered at $q_0=1.2$
with normal deviation $\Delta q_x=0.25$ (shaded region in
\subfigref{Band-diagram}{c}). The resulting pulse profile is localized to
roughly 50 unit cells.

The subsequent evolution of the sound pulse is shown in \figref{wave-packet}
which tracks the instantaneous position along the chain as a function of
time. Video 1 demonstrates the phenomenon in a full-wave finite element 
simulation of the continuum model. 
During the evolution, the dispersion is dynamically 
tuned at two separate intervals (shaded regions in \figref{wave-packet}) by varying $T$
in the continuum model and the parameters $\{\tau,k_1\}$ in the discrete model.
Outside these intervals, the
effective tension $T$ is maintained at the constant values depicted in
\subfigref{Band-diagram}{c}. At $T=120$, the wavepacket has a positive group
velocity (positive slope of the dispersion relation at $q_0$), and travels to
the right (\figref{wave-packet}, top). The tension was then reduced over roughly
$600$ oscillation cycles towards the value
$T = T_\text{fb}$ with zero group velocity (upper shaded region in
\figref{wave-packet}). The slow modulation arrests the pulse, which oscillates
in place with minimal distortion when the tension is maintained at $T_\text{fb}$
(\figref{wave-packet}, middle). Finally, the tension was reduced
further to $T=1$ at which the wavepacket group velocity is negative
(\subfigref{Band-diagram}{c}, bottom panel); this change reverses the
propagation direction of the pulse (\figref{wave-packet}, bottom).

To approximate unavoidable losses in real systems, the simulations reported in
\figref{wave-packet} included linear drag forces which recreate a quality factor (Q-factor) of
$Q \approx 6500$ (Appendix \ref{appendixDynamicalSimulation}).
 While the energy density of the pulse decays as a result, the
damping does not significantly distort the pulse profile or interfere
with the expected zero group velocity (in contrast to metamaterials which
attempt to generate a vanishing group velocity via local
resonances~\cite{Theocharis2014}). Quality factors of order $10^4$ are well
within achievable limits for plate resonators as discussed below. The dynamic tuning could also be executed over fewer oscillations in a
system with lower $Q$, at the cost of increased energy leakage into spurious
modes.

\section{Discussion and conclusion}

\subsection{Considerations for practical realization} 

\paragraph{Tunability and losses.} Experimental implementations of our proposal
would require  plate resonator arrays with high quality factors
and tunable in-plane tension. Micromechanical systems based on two-dimensional
materials are a prime candidate~\cite{Cha2018,graphene-array}. The membranes are
suspended over voids fabricated in semiconductor substrates
with the desired metamaterial geometry, and the boundaries are restricted by
adhesion of the membrane to the semiconductor at the void boundaries.
Prestress modulation of phonon bands via
electrostatic backgating has been demonstrated in SiN resonator arrays with
$Q \approx 1,700$~\cite{Cha2018}, and Q-factors as high as $10^8$ have been
reported for individual SiN resonators~\cite{doi:10.1063/1.4938747}.
Graphene-based resonator arrays~\cite{graphene-array} can be tuned via
electrostatic~\cite{Mei2018} or thermally-induced~\cite{Blaikie2019}
prestresses; Q-factors of order $10^5$ have been
reported~\cite{graphene-Qfactor2,graphene-heterostructure}. 

\paragraph{Disorder.} Another important factor to consider for practical realizations is the effect of 
disorder due to nonuniformities in fabrication. Finite mesh effects introduce a small
amount of spatial disorder in the continuum simulations, which induces energy
leakage from the manipulated wave-packet into spurious modes as can be observed in the lower right corner 
of \figref{wave-packet}(a). This effect can be reproduced  and further
investigated in the discrete model by
deliberately introducing known amounts of quenched disorder in the effective
stiffness parameters. Specifically, we randomly perturb the parameters $\tau$
and $k_1$ to lie within a specified percentage of the
parameter values at $T=1$. The stiffness perturbation on each spring is a random variable
which is kept constant for the duration of the simulation to recreate the
effect of quenched disorder in the fundamental mode frequencies of the
resonators. 
We find that the pulse reversal is robust up to 3\% disorder,
\figref{wave-packet}(b). This amount of precision is well within 
achievable levels for micromechanical systems; for instance,
Ref.~\onlinecite{Cha2018} estimated parameter variations of far less than 1\%
across a nanoelectromechanical resonator array. Furthermore, the fundamental mode frequencies of
membrane-based resonators can be individually corrected after fabrication using
photodoping and electrostatic backgating to achieve the required spatial uniformity~\cite{Miller2020}.

\subsection{Future directions} 

Besides enabling signal reversal through dynamic dispersion tuning, the
physical mechanism reported here also provides a means to realizing 
flat bands. Like their counterparts in electronics~\cite{Derzhko2015} and
photonics~\cite{Leykam2018a}, phononic flat bands~\cite{acoustic-array,PhysRevApplied.8.064031,Matlack2018,metamaterial-lattice} are expected to have
interesting transport, localization, and topological properties. Our design, which is a phononic analog of
recipes for designing isolated electronic~\cite{PhysRevB.96.155137} and
photonic~\cite{Maimaiti2017} flat bands, can be
extended to two-dimensional arrays and combined with
lattice-based strategies to generate additional classes of flat
bands~\cite{PhysRevB.99.045107,PhysRevB.96.155137}. Although we focused on slow
parameter modulation in this work, faster dynamic modulation of prestresses
could enable non-Hermitian and active topological
phenomena~\cite{fleury-paper,PhysRevB.97.014305,Zangeneh-Nejad2019,Scheibner2020,Coulais2021}.
Discoveries of such phenomena could be aided by the analytically-tractable
discrete model which quantitatively reproduces the 
numerically-solved continuum dynamics (\figref{wave-packet}).
Beyond elasticity,
modifying the bonding character of paired degrees of freedom has also been
demonstrated in photonics~\cite{Caselli2012}, quantum dots~\cite{Yakimov2009}, and
superconductors~\cite{Andreev-Molecules}, and has been proposed as a band-tuning
mechanism for ultracold atoms~\cite{PhysRevA.91.023409}.
Our basic approach, summarized in
\figref{illustration}, could be explored in these systems as a route to tunable dispersion
and controlled manipulation of photonic and electronic wavepackets.

\section{Acknowledgments}

We thank Benjam\'in Alem\'an, Andrew Blaikie,
Brittany Carter, and David Miller for inspiration and input on experimental
realizations; and Eric Corwin for useful comments.
We acknowledge support from the College of Arts and
Sciences at the University of Oregon via startup funds to JP.

\appendix

\section{Finite-element analysis of continuum model \label{appendixFiniteElementAnalysis}}

Finite-element analyses were done in the commercially available package COMSOL Multiphysics. 
The \emph{general form pde} module was used to define an eigenvalue problem based on a
fourth order partial differential equation describing thin plate elasticity,
\begin{equation} 
	\nabla \cdot  \bigg[ \left(  u_{xxx} + 2 u_{xyy} - T u_x \right)  \hat{x} + \left(  u_{yyy}  - T u_y \right) \hat{y} \bigg]  =\lambda u,
	\label{FEM}
\end{equation}
where subscripts denote partial derivatives of $u$ with respect to those coordinates.
The Dirichlet boundary condition $u=0$ and Neumann boundary conditions $u_x=u_y=0$ satisfy the clamped 
boundary condition. Simplifying the equation \eqref{FEM} gives
$\nabla^4 u-T\nabla^2 u = \lambda u$, which is the desired eigenvalue problem. 
The simulation methods were tested by comparing numerically-derived eigensolutions 
for the square Laplacian plate (setting $D\rightarrow0$) and circular clamped biharmonic plate (setting
$T\rightarrow0$) to known analytical results. 

\begin{figure}[!htb]
	\noindent\includegraphics[width=\columnwidth]{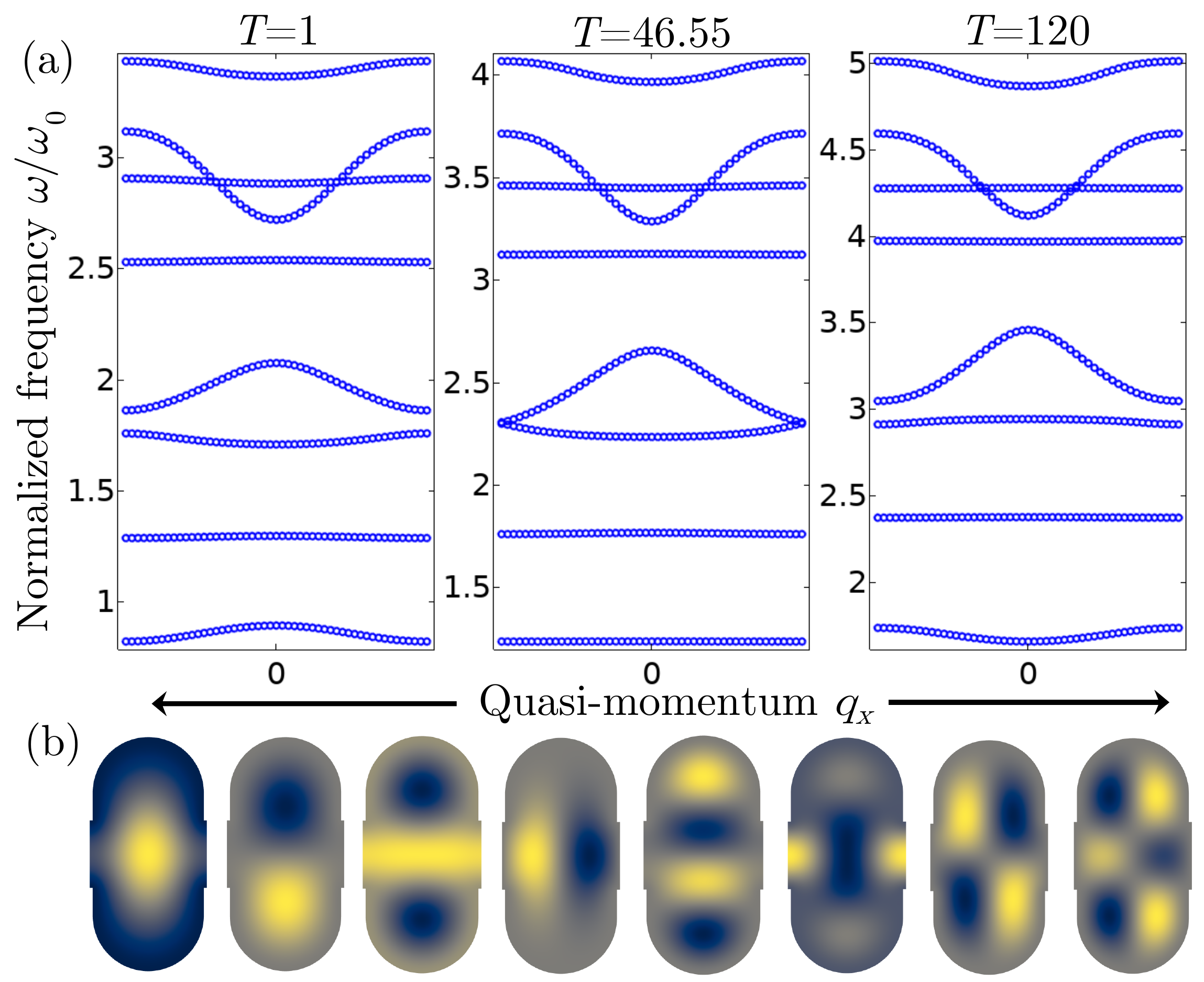}   
	\caption{   
		\label{multiple-bands}
		Eight frequency bands (a) and their eigenfunctions (b) for three different tension $T$ values.
		The change in the dispersive character of the first band intermediated by a flat band can be seen 
		with increasing tension. Displacements in the second and fifth mode (b) can be 
		seen to be isolated from neighboring resonators resulting in mostly non-dispersive 
		bands (a) throughout different tension values. The third and fourth modes have band touching 
		at corners of the Brillouin zone at the flat band tension $T=46.55$ suggesting possible band inversion. 
		The sixth and seventh modes have band crossing and the 
		eight mode simply changes its band width with increasing tension.
	} 
\end{figure}

The continuum thin-plate resonator model has infinitely many bands, of which a
subset are obtained numerically. In \figref{multiple-bands}, the first eight
bands and the eigenfunctions associated with those bands are shown as an
example. In this study, we focused solely on the the lowest band, which is built
primarily from the lowest-frequency, or fundamental, modes of individual
resonators. This is apparent from the mode shape of the Bloch eigenfunction in
\subfigref{multiple-bands}{b}, which mirrors the mode shape of the
single-resonator fundamental mode.

\section{Insensitivity to boundary conditions \label{appendixBoundaryCondition}}

\begin{figure}[!htb]
	\noindent\includegraphics[width=\columnwidth]{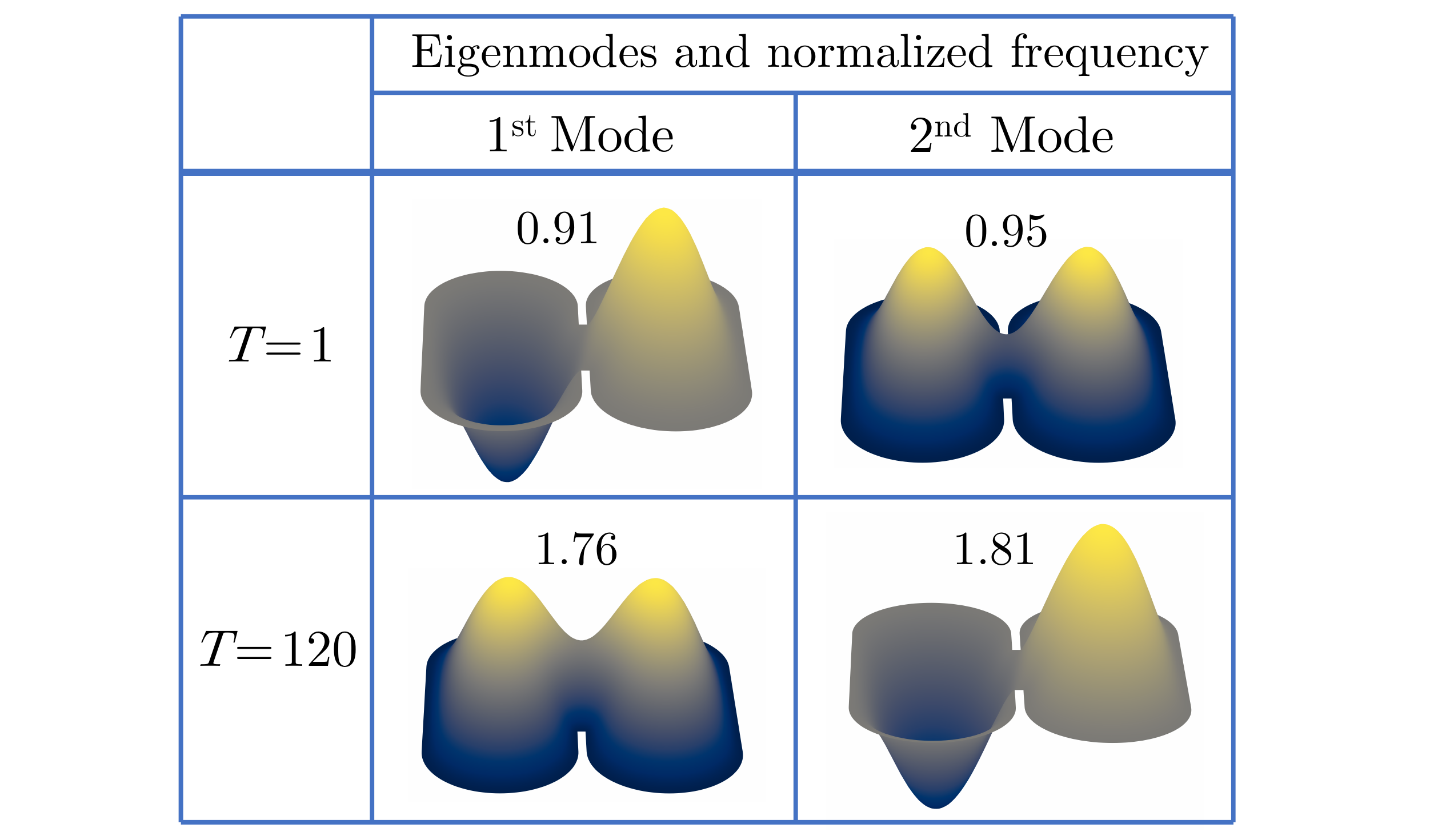}   
	\caption{   
		\label{simply-supported}
		Eigenmodes and normalized frequencies of paired resonators with simply supported boundaries.
		At $T=1$, the system is in the bending dominated region and hence the anti-bonding mode has a lower energy
		and at $T=120$ the system is in the tension dominated region where the bonding mode has a lower energy.
	} 
\end{figure}

The bonding and anti-bonding mode flipping mechanism is insentitive to whether
the domain boundary is clamped or simply supported (Dirichlet boundary condition only). 
In \figref{simply-supported}, the flipping mechanism is shown for the simply supported 
case for two different tension values, $T=1$ and $T=120$. 

\par Having a junction between the resonators is more important than details of the boundary 
condition for the eigenmode switching. A clamped boundary condition was used in this study since 
that is the norm in experimental studies of membrane resonators.

\section{Non-dimensionalization of discrete model \label{appendixDiscreteNonDimensionalization}}
The Newtonian mechanics of the spring-mass chain, with all points assigned a
mass $m$, is described by the second order differential equation
\begin{equation}
m\frac{d^2\mathbf{y}}{d\tilde{t}^2}+\beta\frac{d\mathbf{y}}{d\tilde{t}}+\tilde{\mathbf{K}}\mathbf{y}=0,
\end{equation} 
where $\mathbf{y} = \{y_1,y_2,...\}$ is the vector of vertical displacements,
and $\tilde{\mathbf{K}}$ is the stiffness matrix incorporating the effect of the
on-site, tensed, and torsional springs. For the three-site model in Fig.~2 of
the main text, the stiffness matrix reads
\begin{equation}
\tilde{\mathbf{K}} = {\begin{pmatrix} \tilde{k}_1+\frac{\tilde{\tau}}{\ell}+\frac{\tilde{\kappa}}{\ell^2} & -\frac{\tilde{\tau}}{\ell}-\frac{2\tilde{\kappa}}{\ell^2} & \frac{\tilde{\kappa}}{\ell^2} \\ -\frac{\tilde{\tau}}{\ell}-\frac{2\tilde{\kappa}}{\ell^2} & \tilde{k}_2+\frac{2\tilde{\tau}}{\ell}+\frac{4\tilde{\kappa}}{\ell^2} & -\frac{\tilde{\tau}}{\ell}-\frac{2\tilde{\kappa}}{\ell^2} \\ \frac{\tilde{\kappa}}{\ell^2} & -\frac{\tilde{\tau}}{\ell}-\frac{2\tilde{\kappa}}{\ell^2} & \tilde{k}_1+\frac{\tilde{\tau}}{\ell}+\frac{\tilde{\kappa}}{\ell^2} \end{pmatrix}} 
\label{DynamicalMatrix-3M-full}
,
\end{equation}
where $\ell$ is the horizontal spacing between the masses.
To build a discrete model with
dimensionless parameters that can be related to the continuum system, we choose
$\omega_0^{-1}$ and $a$ as time and length units respectively. The distance
between primary on-site degrees of freedom $y_i$ and $y_{i+2}$ is also set to be
$a$, so that $\ell = a/2$. Upon defining the rescaled time $t=\omega_0\tilde{t}$, spring stiffnesses
$k_i=\tilde{k_i}/m\omega_0^2$, the tension $\tau=2\tilde{\tau}/am\omega_0^2$,
and the torsional stiffness $\kappa=4\tilde{\kappa}/a^2m\omega_0^2$, we obtain
the equation
\begin{equation}
\frac{d^2\mathbf{y}}{dt^2}+2\zeta\frac{d\mathbf{y}}{dt}+\mathbf{K} \mathbf{y}=0,
\end{equation}
where $\mathbf{K}$ is the non-dimensionalized stiffness matrix reported in the
main text, and $\zeta = \beta/(m\omega_0)$ is the damping ratio. The
quality factor is given by $Q=1/2\zeta$. All the parameters presented in the
main text for the discrete model without the tilde symbol are
non-dimensionalized versions.

\section{3-site reduced model and the infinite chain \label{appendix3SiteModel}}
\begin{figure}[htb]
	\noindent\includegraphics[width=\columnwidth]{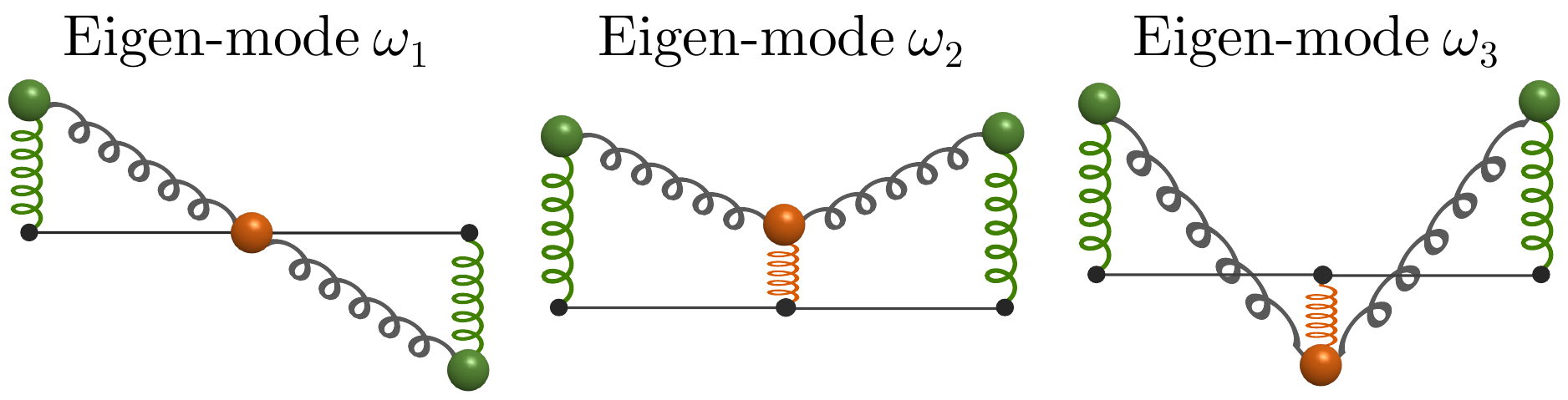}  
	\caption{   
		\label{3-modes}
		Eigenmodes of the 3-site reduced model. In addition to the springs shown in this figure, torsional springs 
		are also attached to the middle masses as shown in the bottom right inset of Fig. 2(c) in the main text.}  
	\vspace{-0.4cm}
\end{figure}

The eigenfrequencies of the 3-site reduced model 
were solved via $|\mathcal{\mathbf{K}}-\omega^2 \mathbf{I}|=0$ with,
\begin{align} 
\begin{split}
\omega_1^2 &=k_1+\tau, \\
\omega_{2,3}^2 &=3\alpha+k_1+k_2 \pm \sqrt{9\alpha^2+(k_2-k_1)(2\alpha+k_2-k_1)},	
\label{3M-eigenvalues}
\end{split}
\end{align}
where, $\alpha=\tau+2\kappa$. The corresponding eigenvectors are visualized in
\figref{3-modes}. The first two modes can be interpreted as antibonding and
bonding modes of the continuum system respectively. The third mode is considerably separated
from the first two modes in frequency for the fit parameters obtained in the
main text, and is not considered in this study.

\paragraph{Infinite chain.} The stiffness matrix for an infinite 1D-chain of the
discrete system takes the form
\begin{equation}
\small
\mathbf{K}_\infty = {\begin{pmatrix}  \vdots &  & \ddots &  & & &   \vdots \\ \dots & \kappa & -\tau-2\kappa & V_1 & -\tau-2\kappa  & \kappa & \dots \\  \dots & 0 & 0 & -\tau-2\kappa & V_2  & -\tau-2\kappa & \dots \\ \vdots & & & & & \ddots & \vdots \end{pmatrix}},
\end{equation}
where, $V_1=k_1+2\tau+2\kappa$ and $V_2=k_2+2\tau+4\kappa$ are on-site spring stiffnesses. The next nearest coupling between the sites with spring stiffness $k_1$ arise due to the torsional spring at the junction. The fourier tranformed version of this infinite matrix is shown in equation 3 of the main text.

\section{Fitting procedure for reduced model parameters \label{appendixFitProcedure}}

Given our choice of physical units, the dimensionless parameters for the
discrete model are fixed by fitting the eigenfrequencies of the discrete model
to the dimensionless frequencies $\omega/\omega_0$ from the continuum
finite-element analysis.

\paragraph{Three-site discrete model.} For the analysis of the resonator pair
(Fig.~2), we aimed to recover the change in frequency of the two lowest modes
upon varying the prestress $T$ in the continuum model by varying the tension
$\tau$ in the reduced model, keeping all other dimensionless parameters fixed.
While $T$ and $\tau$ are related, they are different physical quantities ($T$ is
a force per unit length for the elastic plate, whereas $\tau$ is a tensile force
on the horizontal springs). We assume a simple linear relation $T = c \tau$,
where $c$ is a constant parameter, and find that this relation is sufficient to
recover the mode-crossing behavior.

Given the exact frequencies of the discrete model, \eqnref{3M-eigenvalues}, the
parameter value $k_1=0.8000$ is fixed by equating it to the the square of
antibonding mode frequency from the ontinuum model at $T = \tau = 0$. The
complete relationship between frequency and prestress (tension) for this mode is
then quantitatively recovered by setting $c = 47.5$. Having set these two
parameters, the remaining parameters $k_2=35$ and $\kappa=0.0450$ were fixed by
fitting the analytical form for $\omega_2$ from \eqnref{3M-eigenvalues} to the
bonding mode frequency curve from the continuum model.

\paragraph{Band structure of infinite 1D chain.} 
To obtain quantitative agreement of the discrete model with the continuum
results, both $\tau$ and $k_1$ had to be changed with $T$. Physically, the need
to modify $k_1$ reflects the fact that the bare resonator frequencies themselves
depend on the prestress $T$ in a nontrivial way that depends on geometry. Parameters 
$\kappa\approx0.0575$ and $k_2\approx7.3724$ were fixed across all three
prestress values,  
and parameters $k_1$ and $\tau$ were determined by fitting the analytical
dispersion relation to that from the continuum model at each value of the
prestress $T$. The resulting fit parameters for the three prestress values in
Fig.~3 are: Flat band parameters: $k_1\approx0.5502$ and $\tau\approx0.5217$;
Acoustic-like (top) band parameters: $k_1\approx1.4888$ and $\tau\approx0.7677$; 
Optical-like (bottom) band parameters: $k_1\approx0.0381$ and $\tau\approx0.3195$.

\section{Analytical derivation of flat band in discrete model \label{appendixFlatBandDerivation}}

The existence of a perfectly flat band in the discrete model can be established
provided $\kappa^\prime = 0$ in equation \eqnref{DynamicalMatrix-1D-mod}. We start by assuming arbitrary $\kappa$ and
$\kappa^\prime$, and find the normal mode frequencies of the dynamical matrix
$\mathbf{K}(q_x)$ from equation \eqref{DynamicalMatrix-1D-mod} via
$|\mathbf{K}(q_x) - \omega^2 \mathbf{I}| =0$. This results in the quartic equation,
\begin{align}
\begin{split}
\label{eq:detomega}
(\omega^2)^2 -  & \omega^2  (a+ b \cos q_x+d+e \cos q_x) \\
& +(a + b \cos q_x)(d+e \cos q_x) - 2c^2(1+\cos q_x) = 0
.
\end{split}
\end{align}
The coefficients of $\omega^2$ and $\omega^0 = 1$ are minus the sum and the
product of the two roots $\omega_1^2$ and $\omega_2^2$ respectively.
If we require one of the roots to be a flat band, $\omega_1^2 = \alpha$ for some
constant, then the other root must have the form $$\omega_2^2 =  \gamma + \delta
\cos q_x +\phi \cos^2 q_x$$ to generate the requisite terms in the sum and product. 

By matching coefficients of the $\cos q_x$ term and the remaining term in the sum
and the product of the roots, we can find the relations between constants $\alpha$, $\gamma$, $\delta$, $\phi$, and 
$a$, $b$, $c$, $d$, $e$, and eventually the spring stiffnesses $k_1$, $k_2$,
$\tau$, $\kappa$, $\kappa'$. We immediately find that one of $b$ and $e$, i.e.
one of the two bending stiffnesses, must be zero for the perfectly flat band to
exist. Upon setting $\kappa^\prime$ to zero, the remaining parameters provide
the band dispersion relations
\begin{align}
\begin{split}
\omega_1^2 &= 2\tau + k_1 \\
\omega_2^2 &=  2\tau+ k_2 +6 \kappa  + 2 \kappa \cos q_x \\
\end{split}
\end{align}
with the constraint
\begin{equation}
\label{eq:flatcons}
\tau = -2\kappa + \sqrt{(k_2-k_1)\kappa + (2\kappa)^2}.
\end{equation}
Note that this solution requires $k_2 > k_1$. However, as long as this condition
is fulfilled, a tension $\tau$ can always be found to make the lower band completely
flat.
\par The flat band is used to fit the band related to the fundamental mode 
from the continuum model at $T_\text{fb}=46.55$. For tension values other than $T_\text{fb}$, the band 
is dispersive and must be fitted using the solution to the quartic equation \eqnref{eq:detomega}.

\section{Deviation from perfectly flat band and additional bending stiffness \label{appendixFlatnessDeviation}}

\begin{figure}[htb]
	\vspace{0.5cm}
	\noindent\includegraphics[width=0.9\columnwidth]{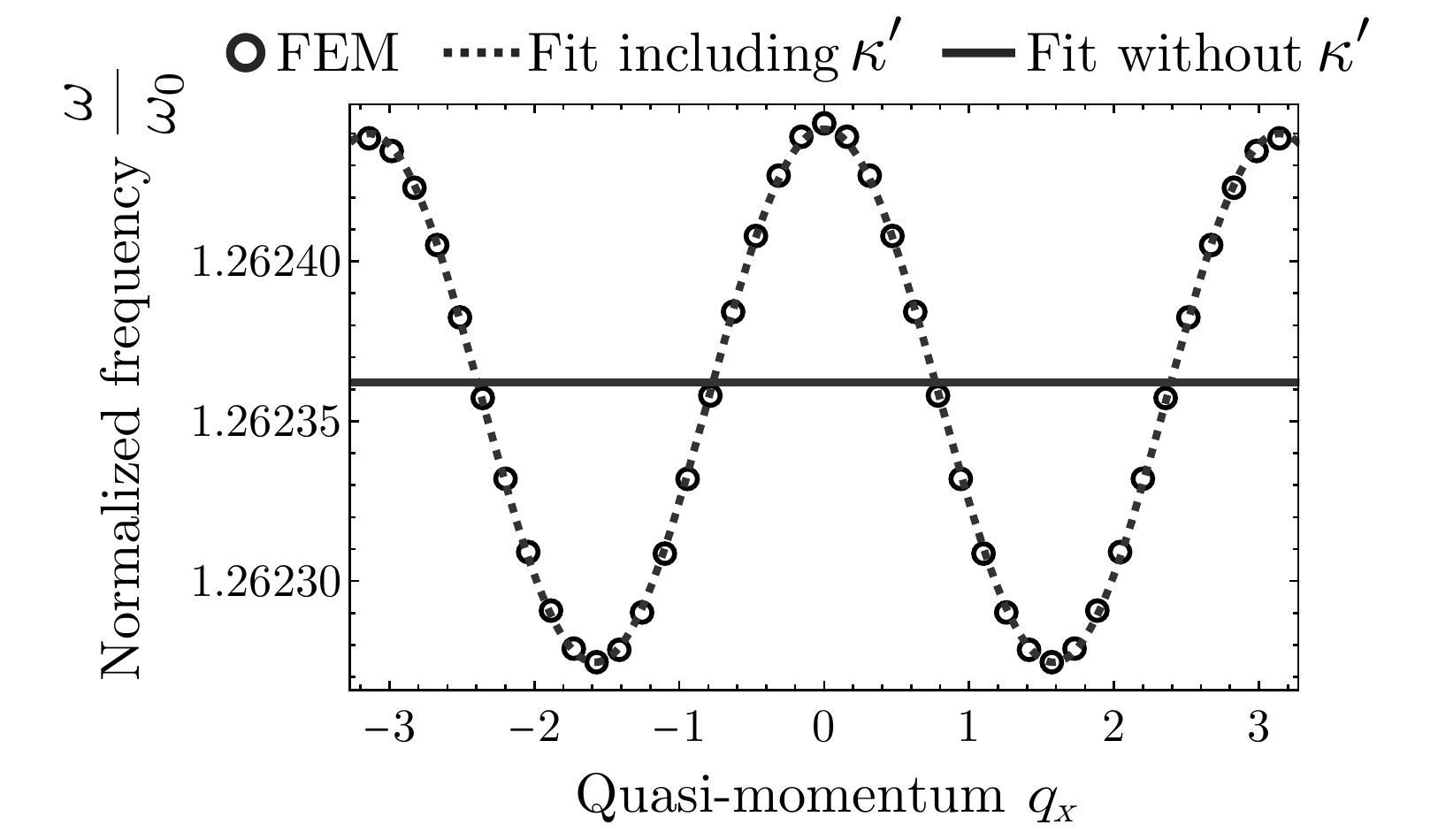}  
	\caption{   
		\label{Adding-second-bending-stiffness}
		Minute deviation from complete flatness. Fit functions are shown for the spring-mass model
		with bending stiffness $\kappa$ only (solid line) at the junction point with the vertical spring stiffness $k_2$ and 
		with additional bending stiffness $\kappa'$ (dotted line) at the point with resonator stiffness $k_1$.}  
	\vspace{-0.5cm}
\end{figure}

At the rescaled tension $T_\text{fb}$, the continuum model of the thin-plate
resonator has a minute deviation from a perfectly flat band with bandwidth that is
$10^{-3}$\% of the mean band frequency (symbols in
\figref{Adding-second-bending-stiffness}).
For practical reasons and for
demonstrating the tunable dispersion character of the fundamental band, such a
small deviation is not consequential. However, even this variation can be
incorporated in the discrete model by including an additional torsional spring,
with rescaled stiffness $\kappa^\prime$,
centered on the resonator degree of freedom (green mass-springs in schematics). 
This addition modifies the equation 3 from main text as follows,
\begin{equation}
	\mathcal{\mathbf{K}}(q_x)= \begin{pmatrix} a+b\cos q_x & -c(1+e^{-iq_x}) \\
		-c(1+e^{iq_x}) & d+e \cos q_x \end{pmatrix}
	\label{DynamicalMatrix-1D-mod}
\end{equation}
where $a=k_1+2\tau+2\kappa+4\kappa'$, $b=2\kappa$, $c=\tau+2\kappa+2\kappa'$,
$d=k_2+2\tau+2\kappa'+4\kappa$, and $e=2\kappa'$.

Upon performing a fit with the additional parameter $\kappa^\prime$, the
deviation from the perfectly flat band is quantitatively recovered (dotted line
in \figref{Adding-second-bending-stiffness}).

\section{Dynamical simulations \label{appendixDynamicalSimulation}}

\textit{Full-wave dynamical simulation} was performed via finite element method in COMSOL for the following 
partial differential equation,
\begin{equation}
	\frac{\partial^2 u}{\partial t^2}+2\gamma \frac{\partial u}{\partial t}+\nabla^4 u-T\nabla^2 u=0. 
      \end{equation}
The second term in the above equation is added to the non-dimensionalized
equation of motion, \eqnref{non-dim-pde}, to incorporate the effect of
dissipation in the continuum simulations. The damping ratio $\gamma$ is chosen
to reproduce the desired quality factor in the fundamental-mode dynamics of a
single resonator. In our study, we used $\gamma\approx8\times10^{-5}$, corresponding 
to a Q-factor $Q \approx 6500$.
A predefined mesh in COMSOL called ``Normal" calibrated for general physics was used for the full-wave 
dynamical simulation. The gradient operators and boundary conditions were
implemented following the steps described in \appref{appendixFiniteElementAnalysis}. Element size parameters for this mesh are as follows: maximum element size = 6.7,
minimum element size = 0.03, maximum element growth rate = 1.3, curvature factor = 0.3, and resolution of 
narrow regions = 1. A time-dependent solver called the generalized alpha was used with a time-step of 0.01. 

\par \textit{Classical dynamical simulation} of the discrete 1D spring-mass model was implemented in C++
using the velocity Verlet algorithm. The position and 
velocity were evolved with time iteration $i$ as follows, 
\begin{align}
	\begin{split}
		\mathbf{y}_{i+1} &\plusequals \mathbf{v}_{i} \Delta t + \frac{1}{2}\mathbf{a}_{i} (\Delta t)^2 \\
		\mathbf{v}_{i+1} &\plusequals \frac{1}{2}(\mathbf{a}_{i+1}+\mathbf{a}_{i}) \Delta t,
		\label{velocity-verlet}
	\end{split}
\end{align}
where, $\plusequals$ is the increment operator in C language and 
$\mathbf{y}=\{y_1,y_2,... ,y_n\}$, $\mathbf{v}$, and $\mathbf{a}$ are the list of particle 
displacements in vertical direction, velocities, and accelerations respectively. The acceleration $a_n$ at 
lattice site $n$ is 
\begin{align}
	\begin{split}
		a_n = \frac{1}{m} \bigg[ & (\tau+2\kappa)(y_{n+1}+y_{n-1}) - \kappa(y_{n+2}+y_{n-2}) \\
		& - (k_1+2\tau+2\kappa) y_n -2\zeta v_n \bigg], \, \text{for even} \, n \\
		\text{and,} \,  a_n = \frac{1}{m} \bigg[ & (\tau+2\kappa)(y_{n+1}+y_{n-1})  \\
		& - (k_2+4\tau+8\kappa) y_n -2\zeta v_n \bigg], \, \text{for odd} \, n, \\
		\label{acceleration}
	\end{split}
\end{align}
based on the dynamical matrix in the Eq. (3) of the main text.
A time-step $\Delta t=0.01$ and a damping ratio $\zeta\approx8\times10^{-5}$ were 
used for the simulation. 


\begin{thebibliography}{52}%
\makeatletter
\providecommand \@ifxundefined [1]{%
 \@ifx{#1\undefined}
}%
\providecommand \@ifnum [1]{%
 \ifnum #1\expandafter \@firstoftwo
 \else \expandafter \@secondoftwo
 \fi
}%
\providecommand \@ifx [1]{%
 \ifx #1\expandafter \@firstoftwo
 \else \expandafter \@secondoftwo
 \fi
}%
\providecommand \natexlab [1]{#1}%
\providecommand \enquote  [1]{``#1''}%
\providecommand \bibnamefont  [1]{#1}%
\providecommand \bibfnamefont [1]{#1}%
\providecommand \citenamefont [1]{#1}%
\providecommand \href@noop [0]{\@secondoftwo}%
\providecommand \href [0]{\begingroup \@sanitize@url \@href}%
\providecommand \@href[1]{\@@startlink{#1}\@@href}%
\providecommand \@@href[1]{\endgroup#1\@@endlink}%
\providecommand \@sanitize@url [0]{\catcode `\\12\catcode `\$12\catcode
  `\&12\catcode `\#12\catcode `\^12\catcode `\_12\catcode `\%12\relax}%
\providecommand \@@startlink[1]{}%
\providecommand \@@endlink[0]{}%
\providecommand \url  [0]{\begingroup\@sanitize@url \@url }%
\providecommand \@url [1]{\endgroup\@href {#1}{\urlprefix }}%
\providecommand \urlprefix  [0]{URL }%
\providecommand \Eprint [0]{\href }%
\providecommand \doibase [0]{https://doi.org/}%
\providecommand \selectlanguage [0]{\@gobble}%
\providecommand \bibinfo  [0]{\@secondoftwo}%
\providecommand \bibfield  [0]{\@secondoftwo}%
\providecommand \translation [1]{[#1]}%
\providecommand \BibitemOpen [0]{}%
\providecommand \bibitemStop [0]{}%
\providecommand \bibitemNoStop [0]{.\EOS\space}%
\providecommand \EOS [0]{\spacefactor3000\relax}%
\providecommand \BibitemShut  [1]{\csname bibitem#1\endcsname}%
\let\auto@bib@innerbib\@empty
\bibitem [{\citenamefont {Wang}\ \emph {et~al.}(2020)\citenamefont {Wang},
  \citenamefont {Wang}, \citenamefont {Wu}, \citenamefont {Chen},\ and\
  \citenamefont {Wang}}]{Wang2020}%
  \BibitemOpen
  \bibfield  {author} {\bibinfo {author} {\bibfnamefont {Y.-F.}\ \bibnamefont
  {Wang}}, \bibinfo {author} {\bibfnamefont {Y.-Z.}\ \bibnamefont {Wang}},
  \bibinfo {author} {\bibfnamefont {B.}~\bibnamefont {Wu}}, \bibinfo {author}
  {\bibfnamefont {W.}~\bibnamefont {Chen}},\ and\ \bibinfo {author}
  {\bibfnamefont {Y.-S.}\ \bibnamefont {Wang}},\ }\bibfield  {title}
  {{\selectlanguage {English}\bibinfo {title} {Tunable and {{Active Phononic
  Crystals}} and {{Metamaterials}}}},\ }\bibfield  {journal} {\bibinfo
  {journal} {Applied Mechanics Reviews}\ }\textbf {\bibinfo {volume} {72}},\
  \href {https://doi.org/10.1115/1.4046222} {10.1115/1.4046222} (\bibinfo
  {year} {2020})\BibitemShut {NoStop}%
\bibitem [{\citenamefont {Wang}\ \emph {et~al.}(2014)\citenamefont {Wang},
  \citenamefont {Casadei}, \citenamefont {Shan}, \citenamefont {Weaver},\ and\
  \citenamefont {Bertoldi}}]{Wang2014}%
  \BibitemOpen
  \bibfield  {author} {\bibinfo {author} {\bibfnamefont {P.}~\bibnamefont
  {Wang}}, \bibinfo {author} {\bibfnamefont {F.}~\bibnamefont {Casadei}},
  \bibinfo {author} {\bibfnamefont {S.}~\bibnamefont {Shan}}, \bibinfo {author}
  {\bibfnamefont {J.~C.}\ \bibnamefont {Weaver}},\ and\ \bibinfo {author}
  {\bibfnamefont {K.}~\bibnamefont {Bertoldi}},\ }\bibfield  {title} {\bibinfo
  {title} {Harnessing {{Buckling}} to {{Design Tunable Locally Resonant
  Acoustic Metamaterials}}},\ }\href
  {https://doi.org/10.1103/PhysRevLett.113.014301} {\bibfield  {journal}
  {\bibinfo  {journal} {Physical Review Letters}\ }\textbf {\bibinfo {volume}
  {113}},\ \bibinfo {pages} {014301} (\bibinfo {year} {2014})}\BibitemShut
  {NoStop}%
\bibitem [{\citenamefont {Bertoldi}(2017)}]{Bertoldi2017a}%
  \BibitemOpen
  \bibfield  {author} {\bibinfo {author} {\bibfnamefont {K.}~\bibnamefont
  {Bertoldi}},\ }\bibfield  {title} {\bibinfo {title} {Harnessing
  {{Instabilities}} to {{Design Tunable Architected Cellular Materials}}},\
  }\href {https://doi.org/10.1146/annurev-matsci-070616-123908} {\bibfield
  {journal} {\bibinfo  {journal} {Annual Review of Materials Research}\
  }\textbf {\bibinfo {volume} {47}},\ \bibinfo {pages} {51} (\bibinfo {year}
  {2017})}\BibitemShut {NoStop}%
\bibitem [{\citenamefont {Babaee}\ \emph {et~al.}(2016)\citenamefont {Babaee},
  \citenamefont {Viard}, \citenamefont {Wang}, \citenamefont {Fang},\ and\
  \citenamefont {Bertoldi}}]{Babaee2016}%
  \BibitemOpen
  \bibfield  {author} {\bibinfo {author} {\bibfnamefont {S.}~\bibnamefont
  {Babaee}}, \bibinfo {author} {\bibfnamefont {N.}~\bibnamefont {Viard}},
  \bibinfo {author} {\bibfnamefont {P.}~\bibnamefont {Wang}}, \bibinfo {author}
  {\bibfnamefont {N.~X.}\ \bibnamefont {Fang}},\ and\ \bibinfo {author}
  {\bibfnamefont {K.}~\bibnamefont {Bertoldi}},\ }\bibfield  {title} {\bibinfo
  {title} {Harnessing {{Deformation}} to {{Switch On}} and {{Off}} the
  {{Propagation}} of {{Sound}}},\ }\href
  {https://doi.org/10.1002/adma.201504469} {\bibfield  {journal} {\bibinfo
  {journal} {Advanced Materials}\ }\textbf {\bibinfo {volume} {28}},\ \bibinfo
  {pages} {1631} (\bibinfo {year} {2016})}\BibitemShut {NoStop}%
\bibitem [{\citenamefont {Hedayatrasa}\ \emph {et~al.}(2016)\citenamefont
  {Hedayatrasa}, \citenamefont {Abhary}, \citenamefont {Uddin},\ and\
  \citenamefont {Guest}}]{Hedayatrasa2016}%
  \BibitemOpen
  \bibfield  {author} {\bibinfo {author} {\bibfnamefont {S.}~\bibnamefont
  {Hedayatrasa}}, \bibinfo {author} {\bibfnamefont {K.}~\bibnamefont {Abhary}},
  \bibinfo {author} {\bibfnamefont {M.~S.}\ \bibnamefont {Uddin}},\ and\
  \bibinfo {author} {\bibfnamefont {J.~K.}\ \bibnamefont {Guest}},\ }\bibfield
  {title} {{\selectlanguage {English}\bibinfo {title} {Optimal design of
  tunable phononic bandgap plates under equibiaxial stretch}},\ }\href
  {https://doi.org/10.1088/0964-1726/25/5/055025} {\bibfield  {journal}
  {\bibinfo  {journal} {Smart Materials and Structures}\ }\textbf {\bibinfo
  {volume} {25}},\ \bibinfo {pages} {055025} (\bibinfo {year}
  {2016})}\BibitemShut {NoStop}%
\bibitem [{\citenamefont {Pal}\ \emph {et~al.}(2016)\citenamefont {Pal},
  \citenamefont {Rimoli},\ and\ \citenamefont {Ruzzene}}]{Pal2016}%
  \BibitemOpen
  \bibfield  {author} {\bibinfo {author} {\bibfnamefont {R.~K.}\ \bibnamefont
  {Pal}}, \bibinfo {author} {\bibfnamefont {J.}~\bibnamefont {Rimoli}},\ and\
  \bibinfo {author} {\bibfnamefont {M.}~\bibnamefont {Ruzzene}},\ }\bibfield
  {title} {{\selectlanguage {English}\bibinfo {title} {Effect of large
  deformation pre-loads on the wave properties of hexagonal lattices}},\ }\href
  {https://doi.org/10.1088/0964-1726/25/5/054010} {\bibfield  {journal}
  {\bibinfo  {journal} {Smart Materials and Structures}\ }\textbf {\bibinfo
  {volume} {25}},\ \bibinfo {pages} {054010} (\bibinfo {year}
  {2016})}\BibitemShut {NoStop}%
\bibitem [{\citenamefont {Casadei}\ \emph {et~al.}(2012)\citenamefont
  {Casadei}, \citenamefont {Delpero}, \citenamefont {Bergamini}, \citenamefont
  {Ermanni},\ and\ \citenamefont {Ruzzene}}]{Casadei2012}%
  \BibitemOpen
  \bibfield  {author} {\bibinfo {author} {\bibfnamefont {F.}~\bibnamefont
  {Casadei}}, \bibinfo {author} {\bibfnamefont {T.}~\bibnamefont {Delpero}},
  \bibinfo {author} {\bibfnamefont {A.}~\bibnamefont {Bergamini}}, \bibinfo
  {author} {\bibfnamefont {P.}~\bibnamefont {Ermanni}},\ and\ \bibinfo {author}
  {\bibfnamefont {M.}~\bibnamefont {Ruzzene}},\ }\bibfield  {title} {\bibinfo
  {title} {Piezoelectric resonator arrays for tunable acoustic waveguides and
  metamaterials},\ }\href {https://doi.org/10.1063/1.4752468} {\bibfield
  {journal} {\bibinfo  {journal} {Journal of Applied Physics}\ }\textbf
  {\bibinfo {volume} {112}},\ \bibinfo {pages} {064902} (\bibinfo {year}
  {2012})}\BibitemShut {NoStop}%
\bibitem [{\citenamefont {Cha}\ and\ \citenamefont {Daraio}(2018)}]{Cha2018}%
  \BibitemOpen
  \bibfield  {author} {\bibinfo {author} {\bibfnamefont {J.}~\bibnamefont
  {Cha}}\ and\ \bibinfo {author} {\bibfnamefont {C.}~\bibnamefont {Daraio}},\
  }\bibfield  {title} {{\selectlanguage {English}\bibinfo {title} {Electrical
  tuning of elastic wave propagation in nanomechanical lattices at {{MHz}}
  frequencies}},\ }\href {https://doi.org/10.1038/s41565-018-0252-6} {\bibfield
   {journal} {\bibinfo  {journal} {Nature Nanotechnology}\ }\textbf {\bibinfo
  {volume} {13}},\ \bibinfo {pages} {1016} (\bibinfo {year}
  {2018})}\BibitemShut {NoStop}%
\bibitem [{\citenamefont {Yi}\ \emph {et~al.}(2019)\citenamefont {Yi},
  \citenamefont {Ouisse}, \citenamefont {{Sadoulet-Reboul}},\ and\
  \citenamefont {Matten}}]{Yi2019}%
  \BibitemOpen
  \bibfield  {author} {\bibinfo {author} {\bibfnamefont {K.}~\bibnamefont
  {Yi}}, \bibinfo {author} {\bibfnamefont {M.}~\bibnamefont {Ouisse}}, \bibinfo
  {author} {\bibfnamefont {E.}~\bibnamefont {{Sadoulet-Reboul}}},\ and\
  \bibinfo {author} {\bibfnamefont {G.}~\bibnamefont {Matten}},\ }\bibfield
  {title} {{\selectlanguage {English}\bibinfo {title} {Active metamaterials
  with broadband controllable stiffness for tunable band gaps and
  non-reciprocal wave propagation}},\ }\href
  {https://doi.org/10.1088/1361-665X/ab19dc} {\bibfield  {journal} {\bibinfo
  {journal} {Smart Materials and Structures}\ }\textbf {\bibinfo {volume}
  {28}},\ \bibinfo {pages} {065025} (\bibinfo {year} {2019})}\BibitemShut
  {NoStop}%
\bibitem [{\citenamefont {Swinteck}\ \emph {et~al.}(2014)\citenamefont
  {Swinteck}, \citenamefont {Lucas},\ and\ \citenamefont
  {Deymier}}]{Swinteck2014}%
  \BibitemOpen
  \bibfield  {author} {\bibinfo {author} {\bibfnamefont {N.}~\bibnamefont
  {Swinteck}}, \bibinfo {author} {\bibfnamefont {P.}~\bibnamefont {Lucas}},\
  and\ \bibinfo {author} {\bibfnamefont {P.~A.}\ \bibnamefont {Deymier}},\
  }\bibfield  {title} {\bibinfo {title} {Optically tunable acoustic wave
  band-pass filter},\ }\href {https://doi.org/10.1063/1.4904075} {\bibfield
  {journal} {\bibinfo  {journal} {AIP Advances}\ }\textbf {\bibinfo {volume}
  {4}},\ \bibinfo {pages} {124603} (\bibinfo {year} {2014})}\BibitemShut
  {NoStop}%
\bibitem [{\citenamefont {Feng}\ and\ \citenamefont {Liu}(2012)}]{Feng2012}%
  \BibitemOpen
  \bibfield  {author} {\bibinfo {author} {\bibfnamefont {R.}~\bibnamefont
  {Feng}}\ and\ \bibinfo {author} {\bibfnamefont {K.}~\bibnamefont {Liu}},\
  }\bibfield  {title} {{\selectlanguage {English}\bibinfo {title} {Tuning the
  band-gap of phononic crystals with an initial stress}},\ }\href
  {https://doi.org/10.1016/j.physb.2012.01.135} {\bibfield  {journal} {\bibinfo
   {journal} {Physica B: Condensed Matter}\ }\textbf {\bibinfo {volume}
  {407}},\ \bibinfo {pages} {2032} (\bibinfo {year} {2012})}\BibitemShut
  {NoStop}%
\bibitem [{\citenamefont {Barnwell}\ \emph {et~al.}(2017)\citenamefont
  {Barnwell}, \citenamefont {Parnell},\ and\ \citenamefont
  {Abrahams}}]{BARNWELL201723}%
  \BibitemOpen
  \bibfield  {author} {\bibinfo {author} {\bibfnamefont {E.~G.}\ \bibnamefont
  {Barnwell}}, \bibinfo {author} {\bibfnamefont {W.~J.}\ \bibnamefont
  {Parnell}},\ and\ \bibinfo {author} {\bibfnamefont {I.~D.}\ \bibnamefont
  {Abrahams}},\ }\bibfield  {title} {\bibinfo {title} {Tunable elastodynamic
  band gaps},\ }\href
  {https://doi.org/https://doi.org/10.1016/j.eml.2016.10.009} {\bibfield
  {journal} {\bibinfo  {journal} {Extreme Mechanics Letters}\ }\textbf
  {\bibinfo {volume} {12}},\ \bibinfo {pages} {23 } (\bibinfo {year}
  {2017})}\BibitemShut {NoStop}%
\bibitem [{\citenamefont {Krushynska}\ \emph {et~al.}(2018)\citenamefont
  {Krushynska}, \citenamefont {Amendola}, \citenamefont {Bosia}, \citenamefont
  {Daraio}, \citenamefont {Pugno},\ and\ \citenamefont
  {Fraternali}}]{Krushynska2018}%
  \BibitemOpen
  \bibfield  {author} {\bibinfo {author} {\bibfnamefont {A.~O.}\ \bibnamefont
  {Krushynska}}, \bibinfo {author} {\bibfnamefont {A.}~\bibnamefont
  {Amendola}}, \bibinfo {author} {\bibfnamefont {F.}~\bibnamefont {Bosia}},
  \bibinfo {author} {\bibfnamefont {C.}~\bibnamefont {Daraio}}, \bibinfo
  {author} {\bibfnamefont {N.~M.}\ \bibnamefont {Pugno}},\ and\ \bibinfo
  {author} {\bibfnamefont {F.}~\bibnamefont {Fraternali}},\ }\bibfield  {title}
  {{\selectlanguage {English}\bibinfo {title} {Accordion-like metamaterials
  with tunable ultra-wide low-frequency band gaps}},\ }\href
  {https://doi.org/10.1088/1367-2630/aad354} {\bibfield  {journal} {\bibinfo
  {journal} {New Journal of Physics}\ }\textbf {\bibinfo {volume} {20}},\
  \bibinfo {pages} {073051} (\bibinfo {year} {2018})}\BibitemShut {NoStop}%
\bibitem [{\citenamefont {Pal}\ \emph {et~al.}(2018)\citenamefont {Pal},
  \citenamefont {Ruzzene},\ and\ \citenamefont {Rimoli}}]{Pal2018a}%
  \BibitemOpen
  \bibfield  {author} {\bibinfo {author} {\bibfnamefont {R.~K.}\ \bibnamefont
  {Pal}}, \bibinfo {author} {\bibfnamefont {M.}~\bibnamefont {Ruzzene}},\ and\
  \bibinfo {author} {\bibfnamefont {J.~J.}\ \bibnamefont {Rimoli}},\ }\bibfield
   {title} {{\selectlanguage {English}\bibinfo {title} {Tunable wave
  propagation by varying prestrain in tensegrity-based periodic media}},\
  }\href {https://doi.org/10.1016/j.eml.2018.06.005} {\bibfield  {journal}
  {\bibinfo  {journal} {Extreme Mechanics Letters}\ }\textbf {\bibinfo {volume}
  {22}},\ \bibinfo {pages} {149} (\bibinfo {year} {2018})}\BibitemShut
  {NoStop}%
\bibitem [{\citenamefont {Li}\ \emph {et~al.}(2020)\citenamefont {Li},
  \citenamefont {Wang},\ and\ \citenamefont {Wang}}]{Li2020a}%
  \BibitemOpen
  \bibfield  {author} {\bibinfo {author} {\bibfnamefont {Z.-N.}\ \bibnamefont
  {Li}}, \bibinfo {author} {\bibfnamefont {Y.-Z.}\ \bibnamefont {Wang}},\ and\
  \bibinfo {author} {\bibfnamefont {Y.-S.}\ \bibnamefont {Wang}},\ }\bibfield
  {title} {{\selectlanguage {English}\bibinfo {title} {Tunable nonreciprocal
  transmission in nonlinear elastic wave metamaterial by initial stresses}},\
  }\href {https://doi.org/10.1016/j.ijsolstr.2019.08.020} {\bibfield  {journal}
  {\bibinfo  {journal} {International Journal of Solids and Structures}\
  }\textbf {\bibinfo {volume} {182-183}},\ \bibinfo {pages} {218} (\bibinfo
  {year} {2020})}\BibitemShut {NoStop}%
\bibitem [{\citenamefont {Fatemi}\ and\ \citenamefont
  {Greenleaf}(1998)}]{Fatemi1998}%
  \BibitemOpen
  \bibfield  {author} {\bibinfo {author} {\bibfnamefont {M.}~\bibnamefont
  {Fatemi}}\ and\ \bibinfo {author} {\bibfnamefont {J.~F.}\ \bibnamefont
  {Greenleaf}},\ }\bibfield  {title} {\bibinfo {title} {Ultrasound-stimulated
  vibro-acoustic spectrography},\ }\href@noop {} {\bibfield  {journal}
  {\bibinfo  {journal} {Science}\ }\textbf {\bibinfo {volume} {280}},\ \bibinfo
  {pages} {82} (\bibinfo {year} {1998})}\BibitemShut {NoStop}%
\bibitem [{\citenamefont {Fu}\ \emph {et~al.}(2017)\citenamefont {Fu},
  \citenamefont {Luo}, \citenamefont {Nguyen}, \citenamefont {Walton},
  \citenamefont {Flewitt}, \citenamefont {Zu}, \citenamefont {Li},
  \citenamefont {McHale}, \citenamefont {Matthews}, \citenamefont {Iborra}
  \emph {et~al.}}]{Fu2017}%
  \BibitemOpen
  \bibfield  {author} {\bibinfo {author} {\bibfnamefont {Y.~Q.}\ \bibnamefont
  {Fu}}, \bibinfo {author} {\bibfnamefont {J.}~\bibnamefont {Luo}}, \bibinfo
  {author} {\bibfnamefont {N.-T.}\ \bibnamefont {Nguyen}}, \bibinfo {author}
  {\bibfnamefont {A.}~\bibnamefont {Walton}}, \bibinfo {author} {\bibfnamefont
  {A.~J.}\ \bibnamefont {Flewitt}}, \bibinfo {author} {\bibfnamefont {X.-T.}\
  \bibnamefont {Zu}}, \bibinfo {author} {\bibfnamefont {Y.}~\bibnamefont {Li}},
  \bibinfo {author} {\bibfnamefont {G.}~\bibnamefont {McHale}}, \bibinfo
  {author} {\bibfnamefont {A.}~\bibnamefont {Matthews}}, \bibinfo {author}
  {\bibfnamefont {E.}~\bibnamefont {Iborra}}, \emph {et~al.},\ }\bibfield
  {title} {\bibinfo {title} {Advances in piezoelectric thin films for acoustic
  biosensors, acoustofluidics and lab-on-chip applications},\ }\href@noop {}
  {\bibfield  {journal} {\bibinfo  {journal} {Progress in Materials Science}\
  }\textbf {\bibinfo {volume} {89}},\ \bibinfo {pages} {31} (\bibinfo {year}
  {2017})}\BibitemShut {NoStop}%
\bibitem [{\citenamefont {Oliner}(1978)}]{1978tap}%
  \BibitemOpen
  \bibinfo {editor} {\bibfnamefont {A.~A.}\ \bibnamefont {Oliner}},\ ed.,\
  \href@noop {} {\emph {\bibinfo {title} {{Acoustic surface waves}}}},\
  \bibinfo {series} {Topics in Applied Physics}, Vol.~\bibinfo {volume} {24}\
  (\bibinfo  {publisher} {Springer},\ \bibinfo {year} {1978})\BibitemShut
  {NoStop}%
\bibitem [{\citenamefont {Li}\ \emph {et~al.}(2014)\citenamefont {Li},
  \citenamefont {Anzel}, \citenamefont {Yang}, \citenamefont {Kevrekidis},\
  and\ \citenamefont {Daraio}}]{Li-Feng}%
  \BibitemOpen
  \bibfield  {author} {\bibinfo {author} {\bibfnamefont {F.}~\bibnamefont
  {Li}}, \bibinfo {author} {\bibfnamefont {P.}~\bibnamefont {Anzel}}, \bibinfo
  {author} {\bibfnamefont {J.}~\bibnamefont {Yang}}, \bibinfo {author}
  {\bibfnamefont {P.~G.}\ \bibnamefont {Kevrekidis}},\ and\ \bibinfo {author}
  {\bibfnamefont {C.}~\bibnamefont {Daraio}},\ }\bibfield  {title} {\bibinfo
  {title} {Granular acoustic switches and logic elements},\ }\href
  {https://doi.org/10.1038/ncomms6311} {\bibfield  {journal} {\bibinfo
  {journal} {Nature Communications}\ }\textbf {\bibinfo {volume} {5}},\
  \bibinfo {pages} {5311} (\bibinfo {year} {2014})}\BibitemShut {NoStop}%
\bibitem [{\citenamefont {{Zangeneh-Nejad}}\ and\ \citenamefont
  {Fleury}(2018)}]{Zangeneh-Nejad2018}%
  \BibitemOpen
  \bibfield  {author} {\bibinfo {author} {\bibfnamefont {F.}~\bibnamefont
  {{Zangeneh-Nejad}}}\ and\ \bibinfo {author} {\bibfnamefont {R.}~\bibnamefont
  {Fleury}},\ }\bibfield  {title} {{\selectlanguage {English}\bibinfo {title}
  {Performing mathematical operations using high-index acoustic
  metamaterials}},\ }\href {https://doi.org/10.1088/1367-2630/aacba1}
  {\bibfield  {journal} {\bibinfo  {journal} {New Journal of Physics}\ }\textbf
  {\bibinfo {volume} {20}},\ \bibinfo {pages} {073001} (\bibinfo {year}
  {2018})}\BibitemShut {NoStop}%
\bibitem [{\citenamefont {Wang}\ \emph {et~al.}(2019)\citenamefont {Wang},
  \citenamefont {Xia}, \citenamefont {Sun}, \citenamefont {Yuan},\ and\
  \citenamefont {Liu}}]{Wang}%
  \BibitemOpen
  \bibfield  {author} {\bibinfo {author} {\bibfnamefont {Y.}~\bibnamefont
  {Wang}}, \bibinfo {author} {\bibfnamefont {J.-p.}\ \bibnamefont {Xia}},
  \bibinfo {author} {\bibfnamefont {H.-x.}\ \bibnamefont {Sun}}, \bibinfo
  {author} {\bibfnamefont {S.-q.}\ \bibnamefont {Yuan}},\ and\ \bibinfo
  {author} {\bibfnamefont {X.-j.}\ \bibnamefont {Liu}},\ }\bibfield  {title}
  {\bibinfo {title} {Binary-phase acoustic passive logic gates},\ }\href
  {https://doi.org/10.1038/s41598-019-44769-0} {\bibfield  {journal} {\bibinfo
  {journal} {Scientific Reports}\ }\textbf {\bibinfo {volume} {9}},\ \bibinfo
  {pages} {8355} (\bibinfo {year} {2019})}\BibitemShut {NoStop}%
\bibitem [{\citenamefont {Caselli}\ \emph {et~al.}(2012)\citenamefont
  {Caselli}, \citenamefont {Intonti}, \citenamefont {Riboli}, \citenamefont
  {Vinattieri}, \citenamefont {Gerace}, \citenamefont {Balet}, \citenamefont
  {Li}, \citenamefont {Francardi}, \citenamefont {Gerardino}, \citenamefont
  {Fiore},\ and\ \citenamefont {Gurioli}}]{Caselli2012}%
  \BibitemOpen
  \bibfield  {author} {\bibinfo {author} {\bibfnamefont {N.}~\bibnamefont
  {Caselli}}, \bibinfo {author} {\bibfnamefont {F.}~\bibnamefont {Intonti}},
  \bibinfo {author} {\bibfnamefont {F.}~\bibnamefont {Riboli}}, \bibinfo
  {author} {\bibfnamefont {A.}~\bibnamefont {Vinattieri}}, \bibinfo {author}
  {\bibfnamefont {D.}~\bibnamefont {Gerace}}, \bibinfo {author} {\bibfnamefont
  {L.}~\bibnamefont {Balet}}, \bibinfo {author} {\bibfnamefont {L.~H.}\
  \bibnamefont {Li}}, \bibinfo {author} {\bibfnamefont {M.}~\bibnamefont
  {Francardi}}, \bibinfo {author} {\bibfnamefont {A.}~\bibnamefont
  {Gerardino}}, \bibinfo {author} {\bibfnamefont {A.}~\bibnamefont {Fiore}},\
  and\ \bibinfo {author} {\bibfnamefont {M.}~\bibnamefont {Gurioli}},\
  }\bibfield  {title} {\bibinfo {title} {Antibonding ground state in photonic
  crystal molecules},\ }\href {https://doi.org/10.1103/PhysRevB.86.035133}
  {\bibfield  {journal} {\bibinfo  {journal} {Physical Review B}\ }\textbf
  {\bibinfo {volume} {86}},\ \bibinfo {pages} {035133} (\bibinfo {year}
  {2012})}\BibitemShut {NoStop}%
\bibitem [{\citenamefont {Yakimov}\ \emph {et~al.}(2009)\citenamefont
  {Yakimov}, \citenamefont {Bloshkin},\ and\ \citenamefont
  {Dvurechenskii}}]{Yakimov2009}%
  \BibitemOpen
  \bibfield  {author} {\bibinfo {author} {\bibfnamefont {A.~I.}\ \bibnamefont
  {Yakimov}}, \bibinfo {author} {\bibfnamefont {A.~A.}\ \bibnamefont
  {Bloshkin}},\ and\ \bibinfo {author} {\bibfnamefont {A.~V.}\ \bibnamefont
  {Dvurechenskii}},\ }\bibfield  {title} {{\selectlanguage {English}\bibinfo
  {title} {Bonding\textendash antibonding ground-state transition in coupled
  {{Ge}}/{{Si}} quantum dots}},\ }\href
  {https://doi.org/10.1088/0268-1242/24/9/095002} {\bibfield  {journal}
  {\bibinfo  {journal} {Semiconductor Science and Technology}\ }\textbf
  {\bibinfo {volume} {24}},\ \bibinfo {pages} {095002} (\bibinfo {year}
  {2009})}\BibitemShut {NoStop}%
\bibitem [{\citenamefont {Zampetaki}\ \emph {et~al.}(2015)\citenamefont
  {Zampetaki}, \citenamefont {Stockhofe},\ and\ \citenamefont
  {Schmelcher}}]{PhysRevA.91.023409}%
  \BibitemOpen
  \bibfield  {author} {\bibinfo {author} {\bibfnamefont {A.~V.}\ \bibnamefont
  {Zampetaki}}, \bibinfo {author} {\bibfnamefont {J.}~\bibnamefont
  {Stockhofe}},\ and\ \bibinfo {author} {\bibfnamefont {P.}~\bibnamefont
  {Schmelcher}},\ }\bibfield  {title} {\bibinfo {title} {Degeneracy and
  inversion of band structure for wigner crystals on a closed helix},\ }\href
  {https://doi.org/10.1103/PhysRevA.91.023409} {\bibfield  {journal} {\bibinfo
  {journal} {Phys. Rev. A}\ }\textbf {\bibinfo {volume} {91}},\ \bibinfo
  {pages} {023409} (\bibinfo {year} {2015})}\BibitemShut {NoStop}%
\bibitem [{\citenamefont {Timoshenko}\ and\ \citenamefont
  {{Woinowsky-Krieger}}(1959)}]{timoshenko1959theory}%
  \BibitemOpen
  \bibfield  {author} {\bibinfo {author} {\bibfnamefont {S.}~\bibnamefont
  {Timoshenko}}\ and\ \bibinfo {author} {\bibfnamefont {S.}~\bibnamefont
  {{Woinowsky-Krieger}}},\ }\href@noop {} {\emph {\bibinfo {title} {Theory of
  Plates and Shells}}},\ Engineering Mechanics Series\ (\bibinfo  {publisher}
  {{McGraw-Hill}},\ \bibinfo {year} {1959})\BibitemShut {NoStop}%
\bibitem [{\citenamefont {Blaikie}\ \emph {et~al.}(2019)\citenamefont
  {Blaikie}, \citenamefont {Miller},\ and\ \citenamefont
  {Alem{\'a}n}}]{Blaikie2019}%
  \BibitemOpen
  \bibfield  {author} {\bibinfo {author} {\bibfnamefont {A.}~\bibnamefont
  {Blaikie}}, \bibinfo {author} {\bibfnamefont {D.}~\bibnamefont {Miller}},\
  and\ \bibinfo {author} {\bibfnamefont {B.~J.}\ \bibnamefont {Alem{\'a}n}},\
  }\bibfield  {title} {{\selectlanguage {English}\bibinfo {title} {A fast and
  sensitive room-temperature graphene nanomechanical bolometer}},\ }\href
  {https://doi.org/10.1038/s41467-019-12562-2} {\bibfield  {journal} {\bibinfo
  {journal} {Nature Communications}\ }\textbf {\bibinfo {volume} {10}},\
  \bibinfo {pages} {1} (\bibinfo {year} {2019})}\BibitemShut {NoStop}%
\bibitem [{\citenamefont {Yosibash}\ and\ \citenamefont
  {Kirby}(2005)}]{Yosibash2005}%
  \BibitemOpen
  \bibfield  {author} {\bibinfo {author} {\bibfnamefont {Z.}~\bibnamefont
  {Yosibash}}\ and\ \bibinfo {author} {\bibfnamefont {R.}~\bibnamefont
  {Kirby}},\ }\bibfield  {title} {{\selectlanguage {English}\bibinfo {title}
  {Dynamic response of various von-kármán non-linear plate models and their
  3-d counterparts}},\ }\href {https://doi.org/10.1016/j.ijsolstr.2004.10.006}
  {\bibfield  {journal} {\bibinfo  {journal} {International Journal of Solids
  and Structures}\ }\textbf {\bibinfo {volume} {42}},\ \bibinfo {pages} {2517}
  (\bibinfo {year} {2005})}\BibitemShut {NoStop}%
\bibitem [{\citenamefont {Sweers}(2001)}]{Sweers2001}%
  \BibitemOpen
  \bibfield  {author} {\bibinfo {author} {\bibfnamefont {G.}~\bibnamefont
  {Sweers}},\ }\bibfield  {title} {{\selectlanguage {English}\bibinfo {title}
  {When is the first eigenfunction for the clamped plate equation of fixed
  sign?}},\ }\href {http://eudml.org/doc/121016} {\bibfield  {journal}
  {\bibinfo  {journal} {Electronic Journal of Differential Equations (EJDE)
  [electronic only]}\ }\textbf {\bibinfo {volume} {2001}},\ \bibinfo {pages}
  {285} (\bibinfo {year} {2001})}\BibitemShut {NoStop}%
\bibitem [{\citenamefont {Brown}\ \emph {et~al.}(1999)\citenamefont {Brown},
  \citenamefont {Davies}, \citenamefont {Jimack},\ and\ \citenamefont
  {Mihajlovi'c}}]{brown1999accurate}%
  \BibitemOpen
  \bibfield  {author} {\bibinfo {author} {\bibfnamefont {B.~M.}\ \bibnamefont
  {Brown}}, \bibinfo {author} {\bibfnamefont {E.~B.}\ \bibnamefont {Davies}},
  \bibinfo {author} {\bibfnamefont {P.~K.}\ \bibnamefont {Jimack}},\ and\
  \bibinfo {author} {\bibfnamefont {M.~D.}\ \bibnamefont {Mihajlovi'c}},\
  }\href@noop {} {\bibinfo {title} {On the accurate finite element solution of
  a class of fourth order eigenvalue problems}} (\bibinfo {year} {1999}),\
  \Eprint {https://arxiv.org/abs/math/9905038} {arXiv:math/9905038 [math.SP]}
  \BibitemShut {NoStop}%
\bibitem [{\citenamefont {Yu}\ \emph {et~al.}(2012)\citenamefont {Yu},
  \citenamefont {Purdy},\ and\ \citenamefont {Regal}}]{Yu2012}%
  \BibitemOpen
  \bibfield  {author} {\bibinfo {author} {\bibfnamefont {P.-L.}\ \bibnamefont
  {Yu}}, \bibinfo {author} {\bibfnamefont {T.~P.}\ \bibnamefont {Purdy}},\ and\
  \bibinfo {author} {\bibfnamefont {C.~A.}\ \bibnamefont {Regal}},\ }\bibfield
  {title} {\bibinfo {title} {Control of {{Material Damping}} in
  {{High}}-\${{Q}}\$ {{Membrane Microresonators}}},\ }\href
  {https://doi.org/10.1103/PhysRevLett.108.083603} {\bibfield  {journal}
  {\bibinfo  {journal} {Physical Review Letters}\ }\textbf {\bibinfo {volume}
  {108}},\ \bibinfo {pages} {083603} (\bibinfo {year} {2012})}\BibitemShut
  {NoStop}%
\bibitem [{\citenamefont {Matlack}\ \emph {et~al.}(2018)\citenamefont
  {Matlack}, \citenamefont {{Serra-Garcia}}, \citenamefont {Palermo},
  \citenamefont {Huber},\ and\ \citenamefont {Daraio}}]{Matlack2018}%
  \BibitemOpen
  \bibfield  {author} {\bibinfo {author} {\bibfnamefont {K.~H.}\ \bibnamefont
  {Matlack}}, \bibinfo {author} {\bibfnamefont {M.}~\bibnamefont
  {{Serra-Garcia}}}, \bibinfo {author} {\bibfnamefont {A.}~\bibnamefont
  {Palermo}}, \bibinfo {author} {\bibfnamefont {S.~D.}\ \bibnamefont {Huber}},\
  and\ \bibinfo {author} {\bibfnamefont {C.}~\bibnamefont {Daraio}},\
  }\bibfield  {title} {{\selectlanguage {English}\bibinfo {title} {Designing
  perturbative metamaterials from discrete models}},\ }\href
  {https://doi.org/10.1038/s41563-017-0003-3} {\bibfield  {journal} {\bibinfo
  {journal} {Nature Materials}\ }\textbf {\bibinfo {volume} {17}},\ \bibinfo
  {pages} {323} (\bibinfo {year} {2018})}\BibitemShut {NoStop}%
\bibitem [{\citenamefont {Theocharis}\ \emph {et~al.}(2014)\citenamefont
  {Theocharis}, \citenamefont {Richoux}, \citenamefont {Garc{\'i}a},
  \citenamefont {Merkel},\ and\ \citenamefont {Tournat}}]{Theocharis2014}%
  \BibitemOpen
  \bibfield  {author} {\bibinfo {author} {\bibfnamefont {G.}~\bibnamefont
  {Theocharis}}, \bibinfo {author} {\bibfnamefont {O.}~\bibnamefont {Richoux}},
  \bibinfo {author} {\bibfnamefont {V.~R.}\ \bibnamefont {Garc{\'i}a}},
  \bibinfo {author} {\bibfnamefont {A.}~\bibnamefont {Merkel}},\ and\ \bibinfo
  {author} {\bibfnamefont {V.}~\bibnamefont {Tournat}},\ }\bibfield  {title}
  {{\selectlanguage {English}\bibinfo {title} {Limits of slow sound propagation
  and transparency in lossy, locally resonant periodic structures}},\ }\href
  {https://doi.org/10.1088/1367-2630/16/9/093017} {\bibfield  {journal}
  {\bibinfo  {journal} {New Journal of Physics}\ }\textbf {\bibinfo {volume}
  {16}},\ \bibinfo {pages} {093017} (\bibinfo {year} {2014})}\BibitemShut
  {NoStop}%
\bibitem [{\citenamefont {Zande}\ \emph {et~al.}(2010)\citenamefont {Zande},
  \citenamefont {Barton}, \citenamefont {Alden}, \citenamefont {Ruiz-Vargas},
  \citenamefont {Whitney}, \citenamefont {Pham}, \citenamefont {Park},
  \citenamefont {Parpia}, \citenamefont {Craighead},\ and\ \citenamefont
  {McEuen}}]{graphene-array}%
  \BibitemOpen
  \bibfield  {author} {\bibinfo {author} {\bibfnamefont {A.~M. v.~d.}\
  \bibnamefont {Zande}}, \bibinfo {author} {\bibfnamefont {R.~A.}\ \bibnamefont
  {Barton}}, \bibinfo {author} {\bibfnamefont {J.~S.}\ \bibnamefont {Alden}},
  \bibinfo {author} {\bibfnamefont {C.~S.}\ \bibnamefont {Ruiz-Vargas}},
  \bibinfo {author} {\bibfnamefont {W.~S.}\ \bibnamefont {Whitney}}, \bibinfo
  {author} {\bibfnamefont {P.~H.~Q.}\ \bibnamefont {Pham}}, \bibinfo {author}
  {\bibfnamefont {J.}~\bibnamefont {Park}}, \bibinfo {author} {\bibfnamefont
  {J.~M.}\ \bibnamefont {Parpia}}, \bibinfo {author} {\bibfnamefont {H.~G.}\
  \bibnamefont {Craighead}},\ and\ \bibinfo {author} {\bibfnamefont {P.~L.}\
  \bibnamefont {McEuen}},\ }\bibfield  {title} {\bibinfo {title} {Large-scale
  arrays of single-layer graphene resonators},\ }\bibfield  {booktitle} {\emph
  {\bibinfo {booktitle} {Nano Letters}},\ }\href
  {https://doi.org/10.1021/nl102713c} {\bibfield  {journal} {\bibinfo
  {journal} {Nano Letters}\ }\textbf {\bibinfo {volume} {10}},\ \bibinfo
  {pages} {4869} (\bibinfo {year} {2010})}\BibitemShut {NoStop}%
\bibitem [{\citenamefont {Yuan}\ \emph {et~al.}(2015)\citenamefont {Yuan},
  \citenamefont {Cohen},\ and\ \citenamefont {Steele}}]{doi:10.1063/1.4938747}%
  \BibitemOpen
  \bibfield  {author} {\bibinfo {author} {\bibfnamefont {M.}~\bibnamefont
  {Yuan}}, \bibinfo {author} {\bibfnamefont {M.~A.}\ \bibnamefont {Cohen}},\
  and\ \bibinfo {author} {\bibfnamefont {G.~A.}\ \bibnamefont {Steele}},\
  }\bibfield  {title} {\bibinfo {title} {Silicon nitride membrane resonators at
  millikelvin temperatures with quality factors exceeding 108},\ }\href
  {https://doi.org/10.1063/1.4938747} {\bibfield  {journal} {\bibinfo
  {journal} {Applied Physics Letters}\ }\textbf {\bibinfo {volume} {107}},\
  \bibinfo {pages} {263501} (\bibinfo {year} {2015})},\ \Eprint
  {https://arxiv.org/abs/https://doi.org/10.1063/1.4938747}
  {https://doi.org/10.1063/1.4938747} \BibitemShut {NoStop}%
\bibitem [{\citenamefont {Mei}\ \emph {et~al.}(2018)\citenamefont {Mei},
  \citenamefont {Lee}, \citenamefont {Xu},\ and\ \citenamefont
  {Feng}}]{Mei2018}%
  \BibitemOpen
  \bibfield  {author} {\bibinfo {author} {\bibfnamefont {T.}~\bibnamefont
  {Mei}}, \bibinfo {author} {\bibfnamefont {J.}~\bibnamefont {Lee}}, \bibinfo
  {author} {\bibfnamefont {Y.}~\bibnamefont {Xu}},\ and\ \bibinfo {author}
  {\bibfnamefont {P.~X.-L.}\ \bibnamefont {Feng}},\ }\bibfield  {title}
  {\bibinfo {title} {Frequency tuning of graphene nanoelectromechanical
  resonators via electrostatic gating},\ }\href@noop {} {\bibfield  {journal}
  {\bibinfo  {journal} {Micromachines}\ }\textbf {\bibinfo {volume} {9}},\
  \bibinfo {pages} {312} (\bibinfo {year} {2018})}\BibitemShut {NoStop}%
\bibitem [{\citenamefont {G{\"u}ttinger}\ \emph {et~al.}(2017)\citenamefont
  {G{\"u}ttinger}, \citenamefont {Noury}, \citenamefont {Weber}, \citenamefont
  {Eriksson}, \citenamefont {Lagoin}, \citenamefont {Moser}, \citenamefont
  {Eichler}, \citenamefont {Wallraff}, \citenamefont {Isacsson},\ and\
  \citenamefont {Bachtold}}]{graphene-Qfactor2}%
  \BibitemOpen
  \bibfield  {author} {\bibinfo {author} {\bibfnamefont {J.}~\bibnamefont
  {G{\"u}ttinger}}, \bibinfo {author} {\bibfnamefont {A.}~\bibnamefont
  {Noury}}, \bibinfo {author} {\bibfnamefont {P.}~\bibnamefont {Weber}},
  \bibinfo {author} {\bibfnamefont {A.~M.}\ \bibnamefont {Eriksson}}, \bibinfo
  {author} {\bibfnamefont {C.}~\bibnamefont {Lagoin}}, \bibinfo {author}
  {\bibfnamefont {J.}~\bibnamefont {Moser}}, \bibinfo {author} {\bibfnamefont
  {C.}~\bibnamefont {Eichler}}, \bibinfo {author} {\bibfnamefont
  {A.}~\bibnamefont {Wallraff}}, \bibinfo {author} {\bibfnamefont
  {A.}~\bibnamefont {Isacsson}},\ and\ \bibinfo {author} {\bibfnamefont
  {A.}~\bibnamefont {Bachtold}},\ }\bibfield  {title} {\bibinfo {title}
  {Energy-dependent path of dissipation in nanomechanical resonators},\ }\href
  {https://doi.org/10.1038/nnano.2017.86} {\bibfield  {journal} {\bibinfo
  {journal} {Nature Nanotechnology}\ }\textbf {\bibinfo {volume} {12}},\
  \bibinfo {pages} {631} (\bibinfo {year} {2017})}\BibitemShut {NoStop}%
\bibitem [{\citenamefont {Will}\ \emph {et~al.}(2017)\citenamefont {Will},
  \citenamefont {Hamer}, \citenamefont {M{\"u}ller}, \citenamefont {Noury},
  \citenamefont {Weber}, \citenamefont {Bachtold}, \citenamefont {Gorbachev},
  \citenamefont {Stampfer},\ and\ \citenamefont
  {G{\"u}ttinger}}]{graphene-heterostructure}%
  \BibitemOpen
  \bibfield  {author} {\bibinfo {author} {\bibfnamefont {M.}~\bibnamefont
  {Will}}, \bibinfo {author} {\bibfnamefont {M.}~\bibnamefont {Hamer}},
  \bibinfo {author} {\bibfnamefont {M.}~\bibnamefont {M{\"u}ller}}, \bibinfo
  {author} {\bibfnamefont {A.}~\bibnamefont {Noury}}, \bibinfo {author}
  {\bibfnamefont {P.}~\bibnamefont {Weber}}, \bibinfo {author} {\bibfnamefont
  {A.}~\bibnamefont {Bachtold}}, \bibinfo {author} {\bibfnamefont {R.~V.}\
  \bibnamefont {Gorbachev}}, \bibinfo {author} {\bibfnamefont {C.}~\bibnamefont
  {Stampfer}},\ and\ \bibinfo {author} {\bibfnamefont {J.}~\bibnamefont
  {G{\"u}ttinger}},\ }\bibfield  {title} {\bibinfo {title} {High quality factor
  graphene-based two-dimensional heterostructure mechanical resonator},\
  }\bibfield  {booktitle} {\emph {\bibinfo {booktitle} {Nano Letters}},\ }\href
  {https://doi.org/10.1021/acs.nanolett.7b01845} {\bibfield  {journal}
  {\bibinfo  {journal} {Nano Letters}\ }\textbf {\bibinfo {volume} {17}},\
  \bibinfo {pages} {5950} (\bibinfo {year} {2017})}\BibitemShut {NoStop}%
\bibitem [{\citenamefont {Miller}\ \emph {et~al.}(2020)\citenamefont {Miller},
  \citenamefont {Blaikie},\ and\ \citenamefont {Alem{\'a}n}}]{Miller2020}%
  \BibitemOpen
  \bibfield  {author} {\bibinfo {author} {\bibfnamefont {D.}~\bibnamefont
  {Miller}}, \bibinfo {author} {\bibfnamefont {A.}~\bibnamefont {Blaikie}},\
  and\ \bibinfo {author} {\bibfnamefont {B.~J.}\ \bibnamefont {Alem{\'a}n}},\
  }\bibfield  {title} {\bibinfo {title} {Nonvolatile {{Rewritable Frequency
  Tuning}} of a {{Nanoelectromechanical Resonator Using Photoinduced
  Doping}}},\ }\href {https://doi.org/10.1021/acs.nanolett.9b05003} {\bibfield
  {journal} {\bibinfo  {journal} {Nano Letters}\ }\textbf {\bibinfo {volume}
  {20}},\ \bibinfo {pages} {2378} (\bibinfo {year} {2020})}\BibitemShut
  {NoStop}%
\bibitem [{\citenamefont {Derzhko}\ \emph {et~al.}(2015)\citenamefont
  {Derzhko}, \citenamefont {Richter},\ and\ \citenamefont
  {Maksymenko}}]{Derzhko2015}%
  \BibitemOpen
  \bibfield  {author} {\bibinfo {author} {\bibfnamefont {O.}~\bibnamefont
  {Derzhko}}, \bibinfo {author} {\bibfnamefont {J.}~\bibnamefont {Richter}},\
  and\ \bibinfo {author} {\bibfnamefont {M.}~\bibnamefont {Maksymenko}},\
  }\bibfield  {title} {\bibinfo {title} {Strongly correlated flat-band systems:
  The route from heisenberg spins to hubbard electrons},\ }\href@noop {}
  {\bibfield  {journal} {\bibinfo  {journal} {International Journal of Modern
  Physics B}\ }\textbf {\bibinfo {volume} {29}},\ \bibinfo {pages} {1530007}
  (\bibinfo {year} {2015})}\BibitemShut {NoStop}%
\bibitem [{\citenamefont {Leykam}\ and\ \citenamefont
  {Flach}(2018)}]{Leykam2018a}%
  \BibitemOpen
  \bibfield  {author} {\bibinfo {author} {\bibfnamefont {D.}~\bibnamefont
  {Leykam}}\ and\ \bibinfo {author} {\bibfnamefont {S.}~\bibnamefont {Flach}},\
  }\bibfield  {title} {\bibinfo {title} {Perspective: {{Photonic}} flatbands},\
  }\href {https://doi.org/10.1063/1.5034365} {\bibfield  {journal} {\bibinfo
  {journal} {APL Photonics}\ }\textbf {\bibinfo {volume} {3}},\ \bibinfo
  {pages} {070901} (\bibinfo {year} {2018})}\BibitemShut {NoStop}%
\bibitem [{\citenamefont {Wu}\ and\ \citenamefont
  {Mei}(2016)}]{acoustic-array}%
  \BibitemOpen
  \bibfield  {author} {\bibinfo {author} {\bibfnamefont {S.}~\bibnamefont
  {Wu}}\ and\ \bibinfo {author} {\bibfnamefont {J.}~\bibnamefont {Mei}},\
  }\bibfield  {title} {\bibinfo {title} {Flat band degeneracy and near-zero
  refractive index materials in acoustic crystals},\ }\href
  {https://doi.org/10.1063/1.4939847} {\bibfield  {journal} {\bibinfo
  {journal} {AIP Advances}\ }\textbf {\bibinfo {volume} {6}},\ \bibinfo {pages}
  {015204} (\bibinfo {year} {2016})},\ \Eprint
  {https://arxiv.org/abs/https://doi.org/10.1063/1.4939847}
  {https://doi.org/10.1063/1.4939847} \BibitemShut {NoStop}%
\bibitem [{\citenamefont {Zhu}\ and\ \citenamefont
  {Semperlotti}(2017)}]{PhysRevApplied.8.064031}%
  \BibitemOpen
  \bibfield  {author} {\bibinfo {author} {\bibfnamefont {H.}~\bibnamefont
  {Zhu}}\ and\ \bibinfo {author} {\bibfnamefont {F.}~\bibnamefont
  {Semperlotti}},\ }\bibfield  {title} {\bibinfo {title} {Double-zero-index
  structural phononic waveguides},\ }\href
  {https://doi.org/10.1103/PhysRevApplied.8.064031} {\bibfield  {journal}
  {\bibinfo  {journal} {Phys. Rev. Applied}\ }\textbf {\bibinfo {volume} {8}},\
  \bibinfo {pages} {064031} (\bibinfo {year} {2017})}\BibitemShut {NoStop}%
\bibitem [{\citenamefont {Li}\ \emph {et~al.}(2019)\citenamefont {Li},
  \citenamefont {Li}, \citenamefont {Christensen},\ and\ \citenamefont
  {Tan}}]{metamaterial-lattice}%
  \BibitemOpen
  \bibfield  {author} {\bibinfo {author} {\bibfnamefont {B.}~\bibnamefont
  {Li}}, \bibinfo {author} {\bibfnamefont {Z.}~\bibnamefont {Li}}, \bibinfo
  {author} {\bibfnamefont {J.}~\bibnamefont {Christensen}},\ and\ \bibinfo
  {author} {\bibfnamefont {K.~T.}\ \bibnamefont {Tan}},\ }\bibfield  {title}
  {\bibinfo {title} {Dual dirac cones in elastic lieb-like lattice
  metamaterials},\ }\href {https://doi.org/10.1063/1.5085782} {\bibfield
  {journal} {\bibinfo  {journal} {Applied Physics Letters}\ }\textbf {\bibinfo
  {volume} {114}},\ \bibinfo {pages} {081906} (\bibinfo {year} {2019})},\
  \Eprint {https://arxiv.org/abs/https://doi.org/10.1063/1.5085782}
  {https://doi.org/10.1063/1.5085782} \BibitemShut {NoStop}%
\bibitem [{\citenamefont {Misumi}\ and\ \citenamefont
  {Aoki}(2017)}]{PhysRevB.96.155137}%
  \BibitemOpen
  \bibfield  {author} {\bibinfo {author} {\bibfnamefont {T.}~\bibnamefont
  {Misumi}}\ and\ \bibinfo {author} {\bibfnamefont {H.}~\bibnamefont {Aoki}},\
  }\bibfield  {title} {\bibinfo {title} {New class of flat-band models on
  tetragonal and hexagonal lattices: Gapped versus crossing flat bands},\
  }\href {https://doi.org/10.1103/PhysRevB.96.155137} {\bibfield  {journal}
  {\bibinfo  {journal} {Phys. Rev. B}\ }\textbf {\bibinfo {volume} {96}},\
  \bibinfo {pages} {155137} (\bibinfo {year} {2017})}\BibitemShut {NoStop}%
\bibitem [{\citenamefont {Maimaiti}\ \emph {et~al.}(2017)\citenamefont
  {Maimaiti}, \citenamefont {Andreanov}, \citenamefont {Park}, \citenamefont
  {Gendelman},\ and\ \citenamefont {Flach}}]{Maimaiti2017}%
  \BibitemOpen
  \bibfield  {author} {\bibinfo {author} {\bibfnamefont {W.}~\bibnamefont
  {Maimaiti}}, \bibinfo {author} {\bibfnamefont {A.}~\bibnamefont {Andreanov}},
  \bibinfo {author} {\bibfnamefont {H.~C.}\ \bibnamefont {Park}}, \bibinfo
  {author} {\bibfnamefont {O.}~\bibnamefont {Gendelman}},\ and\ \bibinfo
  {author} {\bibfnamefont {S.}~\bibnamefont {Flach}},\ }\bibfield  {title}
  {\bibinfo {title} {Compact localized states and flat-band generators in one
  dimension},\ }\href {https://doi.org/10.1103/PhysRevB.95.115135} {\bibfield
  {journal} {\bibinfo  {journal} {Phys. Rev. B}\ }\textbf {\bibinfo {volume}
  {95}},\ \bibinfo {pages} {115135} (\bibinfo {year} {2017})}\BibitemShut
  {NoStop}%
\bibitem [{\citenamefont {Rhim}\ and\ \citenamefont
  {Yang}(2019)}]{PhysRevB.99.045107}%
  \BibitemOpen
  \bibfield  {author} {\bibinfo {author} {\bibfnamefont {J.-W.}\ \bibnamefont
  {Rhim}}\ and\ \bibinfo {author} {\bibfnamefont {B.-J.}\ \bibnamefont
  {Yang}},\ }\bibfield  {title} {\bibinfo {title} {Classification of flat bands
  according to the band-crossing singularity of bloch wave functions},\ }\href
  {https://doi.org/10.1103/PhysRevB.99.045107} {\bibfield  {journal} {\bibinfo
  {journal} {Phys. Rev. B}\ }\textbf {\bibinfo {volume} {99}},\ \bibinfo
  {pages} {045107} (\bibinfo {year} {2019})}\BibitemShut {NoStop}%
\bibitem [{\citenamefont {Fleury}\ \emph {et~al.}(2016)\citenamefont {Fleury},
  \citenamefont {Khanikaev},\ and\ \citenamefont {Al{\`u}}}]{fleury-paper}%
  \BibitemOpen
  \bibfield  {author} {\bibinfo {author} {\bibfnamefont {R.}~\bibnamefont
  {Fleury}}, \bibinfo {author} {\bibfnamefont {A.~B.}\ \bibnamefont
  {Khanikaev}},\ and\ \bibinfo {author} {\bibfnamefont {A.}~\bibnamefont
  {Al{\`u}}},\ }\bibfield  {title} {\bibinfo {title} {Floquet topological
  insulators for sound},\ }\href {https://doi.org/10.1038/ncomms11744}
  {\bibfield  {journal} {\bibinfo  {journal} {Nature Communications}\ }\textbf
  {\bibinfo {volume} {7}},\ \bibinfo {pages} {11744} (\bibinfo {year}
  {2016})}\BibitemShut {NoStop}%
\bibitem [{\citenamefont {Nassar}\ \emph {et~al.}(2018)\citenamefont {Nassar},
  \citenamefont {Chen}, \citenamefont {Norris},\ and\ \citenamefont
  {Huang}}]{PhysRevB.97.014305}%
  \BibitemOpen
  \bibfield  {author} {\bibinfo {author} {\bibfnamefont {H.}~\bibnamefont
  {Nassar}}, \bibinfo {author} {\bibfnamefont {H.}~\bibnamefont {Chen}},
  \bibinfo {author} {\bibfnamefont {A.~N.}\ \bibnamefont {Norris}},\ and\
  \bibinfo {author} {\bibfnamefont {G.~L.}\ \bibnamefont {Huang}},\ }\bibfield
  {title} {\bibinfo {title} {Quantization of band tilting in modulated phononic
  crystals},\ }\href {https://doi.org/10.1103/PhysRevB.97.014305} {\bibfield
  {journal} {\bibinfo  {journal} {Phys. Rev. B}\ }\textbf {\bibinfo {volume}
  {97}},\ \bibinfo {pages} {014305} (\bibinfo {year} {2018})}\BibitemShut
  {NoStop}%
\bibitem [{\citenamefont {{Zangeneh-Nejad}}\ and\ \citenamefont
  {Fleury}(2019)}]{Zangeneh-Nejad2019}%
  \BibitemOpen
  \bibfield  {author} {\bibinfo {author} {\bibfnamefont {F.}~\bibnamefont
  {{Zangeneh-Nejad}}}\ and\ \bibinfo {author} {\bibfnamefont {R.}~\bibnamefont
  {Fleury}},\ }\bibfield  {title} {\bibinfo {title} {Active times for acoustic
  metamaterials},\ }\href {https://doi.org/10.1016/j.revip.2019.100031}
  {\bibfield  {journal} {\bibinfo  {journal} {Reviews in Physics}\ }\textbf
  {\bibinfo {volume} {4}},\ \bibinfo {pages} {100031} (\bibinfo {year}
  {2019})}\BibitemShut {NoStop}%
\bibitem [{\citenamefont {Scheibner}\ \emph {et~al.}(2020)\citenamefont
  {Scheibner}, \citenamefont {Irvine},\ and\ \citenamefont
  {Vitelli}}]{Scheibner2020}%
  \BibitemOpen
  \bibfield  {author} {\bibinfo {author} {\bibfnamefont {C.}~\bibnamefont
  {Scheibner}}, \bibinfo {author} {\bibfnamefont {W.~T.~M.}\ \bibnamefont
  {Irvine}},\ and\ \bibinfo {author} {\bibfnamefont {V.}~\bibnamefont
  {Vitelli}},\ }\bibfield  {title} {\bibinfo {title} {Non-{{Hermitian Band
  Topology}} and {{Skin Modes}} in {{Active Elastic Media}}},\ }\href
  {https://doi.org/10.1103/PhysRevLett.125.118001} {\bibfield  {journal}
  {\bibinfo  {journal} {Physical Review Letters}\ }\textbf {\bibinfo {volume}
  {125}},\ \bibinfo {pages} {118001} (\bibinfo {year} {2020})}\BibitemShut
  {NoStop}%
\bibitem [{\citenamefont {Coulais}\ \emph {et~al.}(2021)\citenamefont
  {Coulais}, \citenamefont {Fleury},\ and\ \citenamefont {{van
  Wezel}}}]{Coulais2021}%
  \BibitemOpen
  \bibfield  {author} {\bibinfo {author} {\bibfnamefont {C.}~\bibnamefont
  {Coulais}}, \bibinfo {author} {\bibfnamefont {R.}~\bibnamefont {Fleury}},\
  and\ \bibinfo {author} {\bibfnamefont {J.}~\bibnamefont {{van Wezel}}},\
  }\bibfield  {title} {{\selectlanguage {English}\bibinfo {title} {Topology and
  broken {{Hermiticity}}}},\ }\href
  {https://doi.org/10.1038/s41567-020-01093-z} {\bibfield  {journal} {\bibinfo
  {journal} {Nature Physics}\ }\textbf {\bibinfo {volume} {17}},\ \bibinfo
  {pages} {9} (\bibinfo {year} {2021})}\BibitemShut {NoStop}%
\bibitem [{\citenamefont {Pillet}\ \emph {et~al.}(2019)\citenamefont {Pillet},
  \citenamefont {Benzoni}, \citenamefont {Griesmar}, \citenamefont {Smirr},\
  and\ \citenamefont {Girit}}]{Andreev-Molecules}%
  \BibitemOpen
  \bibfield  {author} {\bibinfo {author} {\bibfnamefont {J.~D.}\ \bibnamefont
  {Pillet}}, \bibinfo {author} {\bibfnamefont {V.}~\bibnamefont {Benzoni}},
  \bibinfo {author} {\bibfnamefont {J.}~\bibnamefont {Griesmar}}, \bibinfo
  {author} {\bibfnamefont {J.~L.}\ \bibnamefont {Smirr}},\ and\ \bibinfo
  {author} {\bibfnamefont {{\c C}.~{\"O}.}\ \bibnamefont {Girit}},\ }\bibfield
  {title} {\bibinfo {title} {Nonlocal josephson effect in andreev molecules},\
  }\bibfield  {booktitle} {\emph {\bibinfo {booktitle} {Nano Letters}},\ }\href
  {https://doi.org/10.1021/acs.nanolett.9b02686} {\bibfield  {journal}
  {\bibinfo  {journal} {Nano Letters}\ }\textbf {\bibinfo {volume} {19}},\
  \bibinfo {pages} {7138} (\bibinfo {year} {2019})}\BibitemShut {NoStop}%
\end{thebibliography}

%

\end{document}